
\baselineskip=14pt
\hsize=6truein
\vsize=8truein
\font\rm=cmr12
\font\sm=cmr10
\font\big=cmr12 at 14pt
\font\bgbf=cmbx12 at 14pt
\font\bigbf=cmbx12 at 14pt

\rightline{ITP-SB 92-11}
\bigbf
\centerline{Physics beyond quasi-particles:}
\centerline{Spectrum and completeness of the 3 state}
\centerline{superintegrable chiral Potts model}
\vskip 2cm
\big
\centerline{Srinandan Dasmahapatra\footnote{$^\natural$}
{\hbox{\rm dasmah@max.physics.sunysb.edu}}}
\centerline{Rinat Kedem\footnote{$^\sharp$}
{\hbox{\rm rinat@max.physics.sunysb.edu}}}
\centerline{ and}
\centerline{Barry M. McCoy\footnote{$^\flat$}
{\hbox{\rm mccoy@dirac.physics.sunysb.edu}}}
\vskip 2cm
\centerline{Institute for Theoretical Physics}
\centerline{State University of New York}
\centerline{Stony Brook, N.Y. 11794}
\vskip 2cm
\centerline{\bigbf Abstract}
\rm
We find the rules which count the energy levels of the 3 state superintegrable
chiral Potts model and demonstrate that these rules are complete. We
then derive the complete spectrum of excitations in the thermodynamic limit
in the massive phase and demonstrate the existence of excitations which do
not have a quasi-particle form. The physics of these excitations is
compared with the BCS superconductivity spectrum and the counting rules are
compared with the closely related $S=1$ XXZ spin chain.
\vfill
\eject

\leftline{\bgbf 1. Introduction}
\vskip .5cm
Many-body condensed matter physics, as opposed to few body nuclear physics,
molecular physics, or quantum chemistry, is concerned with the limiting
case $N \rightarrow \infty$, where $N$ is the number of particles of the
system. There are, of course, a large number
of physical questions
in which we are interested for these systems, such as response functions at
finite temperature, but conceptually the easiest and most primitive question
is the solution of the $N$ body Schroedinger equation in the $N \rightarrow
\infty$ limit. In this study it is surely conceptually easier to consider
the spectrum of eigenvalues before one investigates the more detailed
information contained in the eigenfunctions. Thus the first most
primitive question asked about such systems as the free Bose or Fermi
gases, the Landau theory of the Fermi liquid, and the BCS theory of
superconductivity is how to characterize the eigenvalues $E_n$ of
$$H_N |n\rangle=E_n(N) |n\rangle\eqno(1.1)$$ where $|n\rangle$ stands
for an eigenvector of the Hamiltonian $H_N$.

Now the condition of thermodynamic stability is that for all $n$
$$E_n > -AN\eqno(1.2)$$
where A is a positive constant and indeed, many-body physics is
 only interesting
for those systems where the ground state energy per particle
$$e_0=\lim_{N\rightarrow\infty }E_{GS}/N\eqno(1.3)$$
exists. Thus to consider the spectrum of (1.1) in the $N\rightarrow \infty$
limit we are most interested in
$$\lim_{N \rightarrow\infty}(E_n-E_{GS}).\eqno(1.4)$$

In the examples above of the free gases, Landau theory, and BCS theory
there are an infinite number of levels $E_n$ for which the limit (1.4)
exists.  These are what we call the low-lying energy levels. However
for these and most other translationally invariant systems the much
stronger property holds, that $$\lim_{N \rightarrow \infty}
(E_n-E_{GS})=\sum_{i,rules}e_i(P_i)\eqno(1.5a)$$ and
$$\lim_{N\rightarrow
\infty}(P_n-P_{GS})=\sum_{i,rules}P_i\eqno(1.5b)$$
where $e_i(P)$ is a
set of single particle energy levels, $P$ is a momentum, and
$\sum_{i,rules}$ indicates that the energy levels are additively
composed with a certain set of rules, which embody bosonic or fermionic
restrictions and internal quantum numbers such as spin or isospin.
This additive composition of the order one terms in the energy levels
in the $N \rightarrow \infty$ limit and the relation of energy to
momentum of (1.5) is what we mean by a
``quasi-particle'' spectrum. (The phrase ``quasi-particle'' has many
related usages in many-body physics. In this paper we will use only
definition (1.5))

The property of possessing a quasi-particle spectrum occurs in a very large
number of translationally invariant many-body systems. It occurs in
free gases, is a key ingredient in the Landau theory of the Fermi liquid,
and is an important consequence of  BCS theory. However, regardless of
how ubiquitous its occurrence in many-body physics the quasi-particle
spectrum does not follow from general properties such as hermiticity and
a finite range assumption on the potential alone. Indeed, in  BCS
theory it can be argued that the quasi-particle nature of the spectrum is
almost an accident because there seems to be no {\it a priori} reason that the
energy of an excited pair should be precisely the sum of
the energies of 2 quasi-particles  having opposite momentum. The
question is thus raised of whether there exist many-body
systems whose spectra  fail to have the quasi-particle form.

The first example of a non quasi-particle spectrum was discovered
in the $N$-state superintegrable chiral Potts model. This model was originally
introduced
to study the quantum commensurate-incommensurate phase transition [1] and
as a representation [2] of Onsager's algebra [3]. Its spectrum
was studied by recursion relations [4] and by functional equations [5,6,7]
and this led to the discovery of the non quasi-particle excitations.
Subsequently one of the most important features of the
non quasi-particle   spectrum was demonstrated  to follow from Onsager's
algebra alone [8]. However the complete spectrum has not
been presented in detail because of the lack of a completeness relation that
counts all states.  The purpose of this paper is to present this completeness
relation for the three state case and to discuss the
physics of the non quasi-particle excitations.

We define the model and summarize the formalism needed for the
computation of the spectrum in section 2. In section 3 we present the
rules which count the states, and in the appendix demonstrate that
these rules are complete. These counting rules, indeed, have a very
interesting relation to the $S=1$ XXZ model [9-15] with anisotropy parameter
$\gamma = \pi /3$ which we discuss in section 4. In section 5 we
compute all the low-lying excitations in the massive phase. In particular,
the complete result which includes the non quasi-particle excitations
is given in (5.13). However, even though the non quasi-particle
excitations are a new phenomenon they can be regarded as a generalization
of the excitation spectrum of the BCS system [16]. This interpretation
is discussed in section 6. We conclude in section 7 with a summary
of the new phenomena found in this model.

\vfill
\eject

\vskip 1cm
\leftline{\bgbf 2. Formulation}
\vskip .5cm
The superintegrable chiral Potts chain is the specialization of the general
integrable chiral Potts chain
$$
{\cal H} = - \sum_{j=1}^{M} \sum_{n=1}^{N-1} \left\{
\bar{\alpha}_n (X_j)^n
+ \alpha_n (Z_{j} Z^{\dagger}_{j+1})^{n} \right\}\eqno(2.1)
$$
with
$$ X_{j} = I_{N}\otimes\cdots\otimes X^{jth}\otimes\cdots\otimes I_{N}
\eqno{(2.2a)} $$
and
$$ Z_{j} = I_{N}\otimes\cdots\otimes Z^{jth}\otimes\cdots\otimes I_{N}
\eqno{(2.2b)} $$
where $I_{N}$ is the $N \times N$ identity matrix, the elements of
the $N \times N$ matrices $Z$ and $X$ are:
$$ Z_{l,m} = \delta_{l,m} \omega^{l-1}, \eqno{(2.3a)} $$
$$X_{l,m} = \delta_{l, m+1} \ \ (\hbox{mod}\ N), \eqno{(2.3b)} $$
$$ \omega = e^{2 \pi i/N} \eqno{(2.4)} $$
and
$$ \alpha_{n} = \exp{[i(2n-N)\phi/N]}/\sin{(\pi n/N)}, \eqno{(2.5a)} $$
$$ \bar{\alpha_{n}} =\lambda\exp{[i(2n-N)\bar{\phi}/N]}/\sin{(\pi n/N)},
 \eqno{(2.5b)} $$
$$ \cos{\phi} = \lambda\cos{\bar{\phi}} \eqno{(2.6)} $$
to the special point
$$\phi = \bar{\phi} = \pi/2.\eqno(2.7)$$

The formalism needed to compute the eigenvalues of (2.1) in the
superintegrable case (2.7) has been presented extensively in ref. [7].
We summarize here only the results needed to study the completeness
of the spectrum.  For detailed derivations the reader is
referred to ref. [7].

The chiral Potts chain is solvable because of its relation to the
2 dimensional chiral Potts statistical mechanical model [17-21]
whose transfer matrix is
$$
T_{\{l\},\{l'\}} = \prod_{j=1}^{M} W_{p,q}^{v}(l_{j}-l_{j}')
W_{p,q}^{h}(l_{j}-l_{j+1}')\eqno(2.8)
$$
where the local Boltzmann weights
$$ {{W_{p,q}^{v}(n)}\over{W_{p,q}^{v}(0)}} =
   \prod_{j=1}^{n} \Bigl({{d_{p} b_{q} - a_{p} c_{q} \omega^{j}}\over
   {b_{p} d_{q} - c_{p} a_{q} \omega^{j}}} \Bigr) \eqno{(2.9a)} $$
and
$$ {{W_{p,q}^{h}(n)}\over{W_{p,q}^{h}(0)}} =
   \prod_{j=1}^{n} \Bigl({{\omega a_{p} d_{q} - d_{p} a_{q} \omega^{j}}\over
   {c_{p} b_{q} - b_{p} c_{q} \omega^{j}}} \Bigr). \eqno{(2.9b)} $$
satisfy the star-triangle equations.
Here $a_{p}, b_{p}, c_{p}, d_{p}$ and $a_{q}, b_{q}, c_{q}, d_{q}$ lie
on the generalized elliptic curve
$$ a^{N}+\lambda b^{N} = \lambda' b^{N}, \ \
 \lambda a^{N} + b^{N} = \lambda' c^{N}\eqno(2.10)$$
with
$$\lambda' = (1-\lambda^{2})^{1/2}.\eqno(2.11)$$

These Boltzmann weights depend only on the difference $l_{j}-l_{j}'$
(and satisfy \hfill\break$W^{v,h}(n+N) = W^{v,h}(n)$) and the Hamiltonian (2.1)
is invariant under the global rotation of all spins by $2 \pi/N$.  Thus
the eigenvalues of $T$ and ${\cal H}$ may be classified by the $Z_{N}$
charge eigenvalue $e^{2 \pi i Q/N}$.  We further note that (2.1) is
obtained from (2.8) in the limit $q \rightarrow p$ as
$$
T_{p,q} = 1\{1 + 2 u M\Bigl({{a_{p} c_{p}}\over{b_{p} d_{p}}}\Bigr)^{1-N/2}
\sum_{l=1}^{N-1} \Bigl({{a_{p} c_{p}}\over{b_{p} d_{p}}}\Bigr)^{l-1}
{{1}\over{1-\omega^{-l}}}+ u {\cal H} + O(u^{2})\}\eqno(2.12)$$
and that
$$
e^{2 i \phi/N} = \omega^{1/2} {{a_{p} c_{p}}\over{b_{p} d_{p}}},
\ \ e^{2 i \bar{\phi}/N} = \omega^{1/2} {{a_{p} d_{p}}\over{b_{p} c_{p}}}.
\eqno(2.13)$$

The method of computation of the spectrum of (2.1) for $N=3$ presented
in ref. [7] is based on the fact that $T_{p,q}$ satisfies a functional
equation.  This equation is derived in detail in ref. [22].  For the
superintegrable case (2.7)
$$a_{p} = b_{p}, \ \ c_{p} = d_{p}\eqno(2.14)$$
and
$${{a_{p}}\over{c_{p}}} = \Bigl[{(1-\lambda)\over(1+\lambda)}\Bigr]^{1/2N}.
\eqno(2.15)$$
In this case we explicitly factor $T_{p,q}$ into its zeros and poles using
$$
T_{p,q} = {{(\eta {{a}\over{d}} -1)^{M}}\over
{[(\eta{{a}\over{d}})^{3}-1]^{M}}} T^{N}_{p,q}\eqno(2.16)$$
with
$$\eta = [(1+\lambda)/(1-\lambda)]^{1/6}\eqno(2.17)$$
(where the index $q$ on $a, b, c, d$ is suppressed) and the functional equation
is explicitly written as
$$\eqalign{
T^{N}_{p,q}T^{N}_{p,Rq}T^{N}_{p,R^{2}q}&=
3^{M} e^{-i P} \{({{a b}\over{c d}}\eta^{2}-1)^{M}
({{a b}\over{c d}}\eta^{2} \omega^{2} -1)^{M} T^{N}_{p,q}\cr
&+({{a b}\over{c d}}\eta^{2}\omega^{2}-1)^{M}
({{a b}\over{c d}}\eta^{2}\omega-1)^{M} T^{N}_{p,R^{2}q}\cr
&+({{a b}\over{c d}}\eta^{2}-1)^{M}({{a b}\over{c d}}\eta^{2}\omega-1)^{M}
T^{N}_{p,R^{4}q} \},}\eqno(2.18)$$
where $R$ is the automorphism
$$
R(a,b,c,d) = (b, \omega a, d, c)\eqno(2.19)
$$
and P is the momentum which is derived from $T_{p,q}$ as
$$
 e^{-i P} = \lim_{q\rightarrow p} T_{p,Rq}.\eqno(2.20)
$$

The functional equation (2.18) is solved by the ansatz (note that the sign
of $v_{l}$ here is defined opposite to that of ref. [7])
$$
T^{N}_{p,q} = 3^{M} (\eta{{a}\over{d}})^{P_{a}}(\eta{{b}\over{c}})^{P_{b}}
({{c^{3}}\over{d^{3}}})^{P_{c}} \prod_{l=1}^{m_{p}}\Bigl({{1-\omega v_{l}
\eta^{2} {{a b}\over{c d}}}\over{1-\omega v_{l}}}\Bigr) \prod_{l=1}^{m_{E}}
\Bigl({{1+\lambda}\over{1-\lambda}}\Bigr)^{1/2}
 \Bigl\{{{a^{3}+b^{3}}\over{2 d^{3}}} \pm w_{l} {{(a^{3}-b^{3})}\over
{(1+\lambda)d^{3}}}\Bigr\}\eqno(2.21)
$$
where the $\pm$ signs are chosen independently for each of
the terms in the product, and $P_{a}, P_{b}, P_{c}$ are positive integers.
The numbers $v_{l}$ satisfy
$$
\Bigl({{\omega^{2}-v_{k}}\over{\omega-v_{k}}}\Bigr)^{M} =
-\omega^{-(P_{a} + P_{b})} \prod_{l=1}^{m_{p}} \Bigl({{v_{k}-\omega^{2} v_{l}}
\over{v_{k}-\omega v_{l}}}\Bigr).\eqno(2.22)
$$
The $w_{l}$ are obtained from
$$
w^{2}_{l}= {1\over4} (1-\lambda)^{2} + {{\lambda}\over{1-t_{l}^{3}}}\eqno(2.23)
$$
where the $t_{l}$ are the roots of the polynomial
$$\eqalign{
P(t) &= t^{-(P_{a}+P_{b})} \{ (\omega^{2} t - 1)^{M}(\omega t - 1)^{M}
 \omega^{P_{a}+P_{b}} \prod_{l = 1}^{m_{p}}
 \Bigl({{1-t v_{l}}\over{1-t^{3}v_{l}^{3}}}\Bigr)\cr
&+ (t-1)^{M}(\omega^{2} t -1)^{M} \prod_{l=1}^{m_{p}}
\Bigl({{1-\omega t v_{l}}\over{1-t^{3} v_{l}^{3}}}\Bigr)\cr
&+ (t-1)^{M} (\omega t -1)^{M} \omega^{-(P_{a}+P_{b})}
 \prod_{l=1}^{m_{p}}
\Bigl({{1-\omega^{2} t v_{l}}\over{1-t^{3} v_{l}^{3}}}\Bigr)\}.}
\eqno(2.24)$$
The eigenvalues of $\cal H$ are thus finally obtained as
$$E= A + B \lambda + 6 \sum_{l=1}^{m_{E}} \pm w_{l}\eqno(2.25)$$
where
$$ A= 3(2 P_{c}+m_{E})-2M \eqno{(2.26a)} $$
$$ B=2(P_{b}-P_{a}) -A \eqno{(2.26b)}$$
and the $\pm$ signs are all independently chosen.  In addition the
momentum $P$ is obtained as
$$
e^{-i P} = \omega^{P_{b}} \prod_{l=1}^{m_{p}}\Bigl({{1-\omega^{2}
v_{l}}\over{1-\omega v_{l}}}\Bigr).\eqno(2.27)$$

The ansatz (2.21) gives more possible eigenvalues than
the allowed number of $3^{M}$. The completeness problem is the determination
of those values of $P_{a}, P_{b}, P_{c}, m_{p}, m_{E}$ and those
 possible solutions $v_{l}$ of (2.22) which actually give eigenvalues of
$\cal H$ and $T$.

 It should also be noted that if in (2.22) we set
$$
v_{k} = e^{\lambda_{k}}\eqno(2.28)$$
we obtain
$$
\Bigl({{\sinh{{1\over2}(\lambda_{k}+ {{2 \pi i}\over{3}})}}\over
{\sinh{{1\over2}(\lambda_{k}- {{2 \pi i}\over{3}})}}} \Bigr)^{M} =
\omega^{M-m_{p}-P_{a}-P_{b}} \prod_{l\neq k}^{m_{p}}
{{\sinh{{1\over2}(\lambda_{k}-\lambda_{l} + {{2 \pi i}\over{3}})}}\over
{\sinh{{1\over2}(\lambda_{k}-\lambda_{l}- {{2 \pi i}\over{3}} ) }}}.
\eqno(2.29)
$$
Up to the phase factor in front this equation is the
Bethe's equation for the spin $S=1$, anisotropy $\gamma=\pi/3$ XXZ
chain with $M$ sites and $|S^{z}|=M-m_{p}$ [12-15].  In the
case where $M-m_p-P_a-P_b\equiv 0$ (mod 3) eqn. (2.29)
is identical with the $S=1$ equation with periodic boundary conditions
Thus the study of completeness of the superintegrable chiral Potts model is
closely related to the completeness of this XXZ chain.
\vfill
\eject
\noindent{\bgbf 3. Completeness Rules}
\vskip .5cm
It is a feature common to all studies of eigenvalues which use functional
equations that further information beyond the functional equation is
needed to determine the multiplicities of the eigenvalues (and in particular
to determine whether or not the multiplicity is zero). It is to be expected
that such additional information can be found by studying the eigenvectors
accompanying the eigenvalues. However the production of the eigenvectors,
though of great importance, is also a problem of substantially increased
complexity and in practice it is desirable to obtain the completeness rules in
advance of an eigenvector computation.

Such a study of completeness rules was recently done for the non-chiral limit
$\phi = \bar{\phi} = 0$ of (2.1) [23] by combining
a finite size study of eigenvalues of
chains of size up to M=7 with a generalization of the string hypothesis [24-27]
approach to the counting of solutions of Bethe's ansatz equations. We will
here follow a similar approach for the superintegrable case (2.7). First
we will use the results of a finite size study to obtain a set of rules
for the allowed integers $P_a, P_b, P_c, m_E $ and  $m_p$, then we will use
a finite size study of the $v_l$ to find a set of rules which counts
all eigenvalues. In the appendix we show that these rules derived from
systems of size $M = 3,4,5,6$ sum up to the total number of states $3^M$
for all $M$.
\vskip 1cm
\noindent{\bf A. Rules for $\bf P_a, P_b, P_c, m_E,$ and $\bf m_p$ }
\vskip .5cm
 From (2.26a) we see that instead of specifying $P_c$ it is sufficient to
specify the coefficient $A$ in (2.25). We thus summarize the results of
the finite size study of ref. [7] by presenting in table 1 the allowed
values of $A+B\lambda, P_a,$ and $P_b$. For each set the integers, $m_E$
are either even or odd. This parity is also listed in table 1.
We note that within each set there is a sum rule on the quantity
$$
3 m_E + 2 m_p + P_a + P_b. \eqno(3.1)
$$
The value of this sum rule is also listed in table 1. In addition
we include the value of $M-P_a-P_b-m_p$ which determines the phase
factor in (2.29).
\vskip 1cm
\noindent{\bf B. Rules for $\bf v_l$}
\vskip .5cm
In table 2 we give for $M=3,4,5 $ and $6$ the number of allowed sets
of $v_k$ (denoted by $N$) which correspond to the allowed values of
$P_a, P_b, m_E,$ and $m_p$ given in table 1. From (2.21) each set of
$v_k$ corresponds to $2^{m_E}$ eigenvalues of $T$ so that the $N$
sets give $ 2^{m_E}\times N$ eigenvalues.  The numbers
$N$ are found from table 1 of ref. [7]. These numbers $N$ must count the
number of allowed solutions of (2.22).

To study the solutions of (2.22) we have computed the zeroes of $T$ by means of
the technique of ref. [23] for $M=3,4,5$ and $6$. From that study we find that
even for M=3 there are only 3 classes of solutions:

\noindent{1. real positive solutions (denoted by $v_k^+$)}
$$v_k>0 \eqno(3.2a)$$
\noindent{2. real negative solutions (denoted by $v_k^-$)}
$$v_k<0 \eqno(3.2b)$$
\noindent{3. complex conjugate pair solutions (denoted by $v_k^{2s}$)}
$${\rm Arg} v_k\sim \pm \pi /3. \eqno(3.2c)$$

Following the conventions of ref. [23] we refer to these as positive parity,
negative parity, and (positive parity) 2 strings, and denote the number
of such roots in a given eigenvalue as $m_+, m_-,$ and $m_{2s}$ respectively.
Clearly

$$
m_+ + m_- + 2m_{2s}=m_p. \eqno(3.3)
$$

Following the practice of refs. [12-15] and [23-27] we make these
three classes of
roots explicit by rewriting (2.22) for $v_k^+ $and $v_k^-$ as
$$\eqalign{
\Bigl({{v_{k}^{\pm} \omega -1}\over{v_{k}^{\pm} -\omega}}\Bigr)^{M}&=
(-1)^{m_{p}-1}\omega^{M-P_{a}-P_{b}-m_{p}}\times  \cr
&\prod_{l=1}^{m_{+}} \Bigl({{v_{k}^{\pm} \omega - v_{l}^{+}}\over
                               {v_{l}^{+} \omega - v_{k}^{\pm}}}\Bigr)
\prod_{l=1}^{m_{-}} \Bigl({{v_{k}^{\pm} \omega - v_{l}^{-}}\over
                               {v_{l}^{-} \omega - v_{k}^{\pm}}}\Bigr)\times\cr
&\prod_{l=1}^{m_{2s}} \Bigl[\Bigl({{v_{k}^{\pm} \omega - v_{l}^{2s}}\over
                               {v_{l}^{2s} \omega - v_{k}^{\pm}}}\Bigr)
                            \Bigl({{v_{k}^{\pm} \omega - v_{l}^{2s*}}\over
                             {v_{l}^{2s*} \omega - v_{k}^{\pm}}}\Bigr)\Bigr].}
\eqno(3.4a)
$$
Also, multiplying the equations for $v_{k}^{2s}$ and $v_{k}^{2s*}$
together we obtain
$$
\eqalign{\Bigl[\Bigl({{v_{k}^{2s} \omega -1}\over{v_{k}^{2s}} - \omega}\Bigr)
\Bigl({{v_{k}^{2s*} \omega -1}\over{v_{k}^{2s*}-\omega}}\Bigr)\Bigr]^{M}&=
\omega^{2(M-P_{a}-P_{b}-m_{p})}\times\cr
&\hbox{\hskip-80pt}\prod_{l=1}^{m_{+}}\Bigl[\Bigl({{v_{k}^{2s}\omega-v_{l}^{+}}\over
                                    {v_{l}^{+}\omega-v_{k}^{2s}}}\Bigr)
                         \Bigl({{v_{k}^{2s*}\omega-v_{l}^{+}}\over
                                    {v_{l}^{+}\omega-v_{k}^{2s*}}}\Bigr)
\Bigr]
\prod_{l=1}^{m_{-}}\Bigl[\Bigl({{v_{k}^{2s}\omega-v_{l}^{-}}\over
                                    {v_{l}^{-}\omega-v_{k}^{2s}}}\Bigr)
                         \Bigl({{v_{k}^{2s*}\omega-v_{l}^{-}}\over
                                    {v_{l}^{-}\omega-v_{k}^{2s*}}}\Bigr)
\Bigr]\times\cr
&\hbox{\hskip-80pt}\prod_{l=1}^{m_{2s}}
\Bigl[\Bigl({{v_{k}^{2s}\omega-v_{l}^{2s}}\over
                                     {v_{l}^{2s}\omega-v_{k}^{2s}}}\Bigr)
                          \Bigl({{v_{k}^{2s*}\omega-v_{l}^{2s}}\over
                                     {v_{l}^{2s}\omega-v_{k}^{2s*}}}\Bigr)
                          \Bigl({{v_{k}^{2s}\omega-v_{l}^{2s*}}\over
                                     {v_{l}^{2s*}\omega-v_{k}^{2s}}}\Bigr)
                          \Bigl({{v_{k}^{2s*}\omega-v_{l}^{2s*}}\over
                                     {v_{l}^{2s*}\omega-v_{k}^{2s*}}}\Bigr)
\Bigr] .}\eqno(3.4b)
$$

We continue to follow [12-15] and [23-27] by taking the logarithm of (3.4).
This introduces integers or half integers [28] as the various
branches of $\ln1$
or $\ln(-1)$. These counting integers are of great importance in classifying
the solutions. Their properties are dependent on the choices of branches for
logarithms. The principle used to choose branches is discussed in  reference
[23]. Thus we define all logarithms to obey
$$
-\pi < \hbox{Im}\ln{x} < \pi \ \ (\hbox{and}\ln{1} = 0)\eqno(3.5)
$$
and we use the definitions
$$ i t_{+}(v^{+}) =
\ln\Bigl(-{{v^{+}\omega-1}\over{v^{+}-\omega}}\Bigr),\eqno(3.6a)$$
$$ i t_{-}(v^{-}) = \ln\Bigl({{v^{-}\omega-1}\over{v^{-}-\omega}}\Bigr),
\eqno(3.6b)$$
$$i
t_{2s}(v^{2s})=\ln\Bigl[\Bigl({{v_{k}^{2s}\omega-1}\over{v_{k}^{2s}-\omega}}\Bigr)
\Bigl({{v_{k}^{2s*}\omega-1}\over{v_{k}^{2s*}-\omega}}\Bigr)\Bigr],
\eqno(3.6c)$$
and
$$i \Theta_{+,+}(v_{k}^{+},v_{l}^{+}) = \ln{\Bigl({{v_{k}^{+}\omega
-v_{l}^{+}}\over{v_{l}^{+}\omega-v_{k}^{+}}}\Bigr)},
\eqno(3.7a)$$
$$i \Theta_{+,-}(v_{k}^{+},v_{l}^{-}) = \ln{\Bigl(-{{v_{k}^{+}\omega
-v_{l}^{-}}\over{v_{l}^{-}\omega-v_{k}^{+}}}\Bigr)},
\eqno(3.7b)$$
$$ i \Theta_{-,-}(v_{k}^{-},v_{l}^{-}) = \ln{\Bigl({{v_{k}^{-}\omega
-v_{l}^{-}}\over{v_{l}^{-}\omega-v_{k}^{-}}}\Bigr)},
\eqno(3.7c)$$
$$ i \Theta_{+,2s}(v_{k}^{+},v_{l}^{2s})= \ln{\Bigl[\Bigl({{v_{k}^{+}\omega
-v_{l}^{2s}}\over{v_{l}^{2s}\omega-v_{k}^{+}}}\Bigr)\Bigl({{v_{k}^{+}\omega
-v_{l}^{2s*}}\over{v_{l}^{2s*}\omega-v_{k}^{+}}}\Bigr)\Bigr]},\eqno(3.7d)
$$
$$ i \Theta_{-,2s}(v_{k}^{-},v_{l}^{2s})=\ln{\Bigl[-\Bigl({{v_{k}^{-}\omega
-v_{l}^{2s}}\over{v_{l}^{2s}\omega-v_{k}^{-}}}\Bigr)\Bigl({{v_{k}^{-}\omega
-v_{l}^{2s*}}\over{v_{l}^{2s*}\omega-v_{k}^{-}}}\Bigr)\Bigr]},\eqno(3.7e)
$$
$$\eqalign{i\Theta_{2s,2s}(v_{k}^{2s},v_{l}^{2s})&=
\ln\Bigl[ -\Bigl( {{v_{k}^{2s}\omega-v_{l}^{2s}}
 \over{v_{l}^{2s}\omega-v_{k}^{2s}}}\Bigr)
\Bigl( {{v_{k}^{2s*}\omega-v_{l}^{2s}}
\over{v_{l}^{2s}\omega-v_{k}^{2s*}}}\Bigr)\times\cr
&\Bigl({{v_{k}^{2s}\omega-v_{l}^{2s*}}
\over{v_{l}^{2s*}\omega-v_{k}^{2s}}}\Bigr)
\Bigl({{v_{k}^{2s*}\omega-v_{l}^{2s*}}
\over{v_{l}^{2s*}\omega-v_{k}^{2s*}}}\Bigr)\Bigr]}\eqno(3.7f)
$$
and
$$
\Theta_{m,n}(x,y)=-\Theta_{n,m}(y,x).\eqno(3.8)
$$
We  note in particular that the minus signs in these definitions are chosen
so that
$$
t_{+}(1) = t_{-}(-1) = t_{2s}(-\omega^{\pm 1}) = 0.\eqno(3.9)
$$

Taking the logarithms of (3.4), we define the counting integers $I_{k}^{i}$
by
$$\eqalign{
M t_{+}(v_{k}^{+})&= {1 \over i}\ln{\omega^{M-P_{a}-P_{b}-m_{p}}}
+ 2 \pi I_{k}^{+}\cr
&+ \sum_{l=1}^{m_{+}} \Theta_{+,+}(v_{k}^{+},v_{l}^{+})
+ \sum_{l=1}^{m_{-}}\Theta_{+,-}(v_{k}^{+},v_{l}^{-}) +
\sum_{l=1}^{m_{2s}}\Theta_{+,2s}(v_{k}^{+},v_{l}^{2s}), }\eqno(3.10a)
$$
$$\eqalign{
M t_{-}(v_{k}^{-}) &= {1\over i}\ln{\omega^{M-P_{a}-P_{b}-m_{p}}}
+ 2 \pi I_{k}^{-} +\cr
&\sum_{l=1}^{m_{+}}\Theta_{-,+}(v_{k}^{-},v_{l}^{+}) +
\sum_{l=1}^{m_{-}}\Theta_{-,-}(v_{k}^{-},v_{l}^{-}) +
\sum_{l=1}^{m_{2s}}\Theta_{-,2s}(v_{k}^{-},v_{l}^{2s}),}\eqno(3.10b)
$$
$$
\eqalign{M t_{2s}(v_{k}^{2s}) &= {1 \over
i}\ln{\omega^{2(M-P_{a}-P_{b}-m_{p})}}
+ 2 \pi I_{k}^{2s} +
 \sum_{l=1}^{m_{+}}\Theta_{2s,+}(v_{k}^{2s},v_{l}^{+}) \cr
&\hbox{\hskip30pt}+\sum_{l=1}^{m_{-}}\Theta_{2s,-}(v_{k}^{2s},v_{l}^{-}) +
\sum_{l=1}^{m_{2s}}\Theta_{2s,2s}(v_{k}^{2s},v_{l}^{2s}),}\eqno(3.10c)
$$
where
$$\hbox{$I_k^+$ is} \quad\Bigl\{
 \matrix{\hbox{an integer}\hfill\cr\hbox{a half-integer}\hfill\cr}\Bigr\}
\quad\hbox{if $M+1+m_{+}$
is}\quad\Bigl\{\matrix{\hbox{even}\hfill\cr
\hbox{odd}\hfill\cr}\Bigr\}.\eqno(3.11a)$$
$$\hbox{$I_k^-$ is} \quad\Bigl\{
 \matrix{\hbox{an integer}\hfill\cr\hbox{a half-integer}\hfill\cr}\Bigr\}
\quad\hbox{if $m_{-}+m_{2s}+1$
is}\quad\Bigl\{\matrix{\hbox{even}\hfill\cr
\hbox{odd}\hfill\cr}\Bigr\}.\eqno(3.11b)$$
$$\hbox{$I_k^{2s}$ is} \quad\Bigl\{
 \matrix{\hbox{an integer}\hfill\cr\hbox{a half-integer}\hfill\cr}\Bigr\}
\quad\hbox{if $m_{-}+m_{2s}+1$
is}\quad\Bigl\{\matrix{\hbox{even}\hfill\cr
\hbox{odd}\hfill\cr}\Bigr\}.\eqno(3.11c)$$

We are now able to discuss the rules for completeness of the $v_{l}$.
As will become apparent, the cases $Q=0$, $1$ and $2$ must be treated
separately.

\vskip1cm
\noindent{\bf 1. Q=0}
\vskip .5cm
The completeness rules are simplest for Q=0.  Following the procedure of
ref. [23-25], we count  solutions of (3.10) by finding the
allowed values $I_k^i$ and assuming a monotonic relation between $I_k^i$ and
$v_k^i$. We thus find the limits within which the integers $I_k^i$ can
lie by examining the extreme cases of (3.10), where $|v_k^i|=0$ and $\infty$.
This is easiest for $M\equiv m_p \equiv 0$ (mod 3), $P_a=P_b=0$, where
we find that the number of allowed values for $I_k^+, I_k^-, I_k^{2s}$
are
$$
N_+(m_+,m_-,m_{2s}) = {{{1\over 3}(M-m_p)+m_-}\choose{m_+}},\eqno(3.12a)$$
$$
N_-(m_+,m_-,m_{2s}) = {{{1\over3}(2M+m_p)-m_+-m_{2s}}\choose{m_-}},
\eqno(3.12b)$$
$$
N_{2s}(m_+,m_-,m_{2s}) = {{{1\over3}(2M+m_p)-m_+-m_{2s}}\choose{m_{2s}}},
\eqno(3.12c)$$
where
$$ {{a}\choose{b}} = {{a!}\over{(a-b)! b!}}. \eqno(3.13) $$

Thus the total number of solutions to (3.10) with $I_k^+,\ I_k^-,\ I_k^{2s}$
separately distinct is
$$
N(m_+,m_-,m_{2s}) = N_+ N_- N_{2s}. \eqno(3.14) $$
This counting, however, does not give the correct result.  This can be
seen for the cases $M=6$, $m_p=3$ and 6 where from (3.14) we find the
total number of states satisfying the sum rule (3.3) is 50 and 141,
whereas we see in table 2 that the correct number of states is 46 and 43.
Thus some further correlation among the integers $I_k^i$ is required.
Guided by ref. [23], we make the additional conjecture
$$I_l^{2s} \neq -I_k^-. \eqno(3.15)$$
We compute $I_k^j$ from (3.10) for the solutions $v_k$ obtained from
our $M=6$ finite size study and find that indeed (3.15) is satisfied.
Furthermore, we find that the counting of states given by (3.12) and (3.15)
does indeed give 46 for $m_p = 3$ and 43 for $m_p=6$ as required.

For the case $M\equiv 1,2$ (mod 3) the rules are almost as simple, except
we must use the function $[x]$ which is the greatest integer less than or equal
to $x$.  Thus we obtain the
result valid for $Q=0$ and all $M$, that for all states allowed by table
1, the number of allowed solutions to the Bethe's equations (3.4) for
$v_l$ are

$$\eqalign{N(m_+,m_-,m_{2s}) &= {{\Bigl[{{M+1-m_p}\over{3}}\Bigr]+m_-}\choose
{m_+}}
{{\Bigl[{{2M+m_p}\over{3}}\Bigr]-m_+ -m_{2s}}\choose{m_-}} \times \cr
&{{\Bigl[{{2M+m_p}\over{3}}\Bigr]-m_+ - m_- -
m_{2s}}\choose{m_{2s}}}}\eqno(3.16)
$$
where $I_k^-$ and $I_k^{2s}$ obey (3.15).
\vskip 1cm
\noindent{\bf 2. Q=1}
\vskip .5cm
For Q=1 our procedure is very similar, except now, as may be inferred from
table 1, several separate cases must be considered.
\vskip 1cm
\noindent{\underbar{a. $M\equiv 2$ \ (mod 3), $P_a=P_b=0$ for all $m_p$}}
\vskip .5cm
Here the counting rule is identical with (3.16):
$$\eqalign{N &= {{\Bigl[{{M+1-m_p}\over 3}\Bigr] +m_-}\choose{m_+}}
           {{\Bigl[{{2M+m_p}\over 3}\Bigr] -m_+ - m_{2s}}\choose{m_-}}\times\cr
&\hbox{\hskip30pt}{{\Bigl[{{2M+m_p}\over 3}\Bigr]
-m_+-m_--m_{2s}}\choose{m_{2s}}}}
\eqno(3.17)
$$
\vfill\eject
\vskip 1cm
\leftline{\underbar{b. $M\equiv 0$\ (mod 3), $P_a=2, P_b=0, m_p\equiv 0,1$ (mod
3) or}}
\leftline{\hskip12pt\underbar{$M\equiv1$ (mod 3), $P_a=0, P_b=2, m_p\equiv1, 2$
(mod 3)}}
\vskip .5cm
$$\eqalign{N(m_+,m_-,m_{2s})&=
{{\Bigl[{{M-m_p}\over 3}\Bigr] +m_-}\choose{m_+}}
{{\Bigl[{{2M-2+m_p}\over 3}\Bigr]-m_+-m_{2s}}\choose{m_-}}\times \cr
&\hbox{\hskip30pt}
{{\Bigl[{{2M-2+m_p}\over 3}\Bigr]-m_+-m_--m_{2s}}\choose{m_{2s}}}}\eqno(3.18a)
$$
\vskip 1cm
\leftline{\underbar{c. $M\equiv 0$ (mod 3), $P_a = 0, P_b = 1,
m_p \equiv 0,2$ (mod 3) or}}
\leftline{\hskip12pt\underbar{$M\equiv 1$ (mod 3), $P_a=1, P_b=0, m_p\equiv 0,
1$ (mod 3)}}
\vskip .5cm
$$\eqalign{N(m_+,m_-,m_{2s})&=
{{\Bigl[{{M-m_p+2}\over 3}\Bigr] +m_-}\choose{m_+}}
{{\Bigl[{{2M-2+m_p}\over 3}\Bigr]-m_+-m_{2s}}\choose{m_-}}\times \cr
&\hbox{\hskip30pt}
{{\Bigl[{{2M-2+m_p}\over 3}\Bigr]-m_+-m_--m_{2s}}\choose{m_{2s}}}}\eqno(3.18b)
$$
\vskip 1 cm
\noindent{\bf 3. Q=2}
\vskip .5cm
The rules for Q=2 are obtained by an analogous procedure.  The results are:
\vskip 1 cm
\noindent{\underbar{a. $M\equiv 1$ (mod 3), $P_a=P_b=0$ for all $m_p$}}
\vskip .5cm
$$\eqalign{N(m_+,m_-,m_{2s})&=
{{\Bigl[{{M-m_p+1}\over 3}\Bigr] +m_-}\choose{m_+}}
{{\Bigl[{{2M+m_p}\over 3}\Bigr]-m_+-m_{2s}}\choose{m_-}}\times \cr
&\hbox{\hskip30pt}
{{\Bigl[{{2M+m_p}\over 3}\Bigr]-m_+-m_--m_{2s}}\choose{m_{2s}}}}\eqno(3.19)
$$
\vfill\eject
\vskip 1cm
\leftline{\underbar{b. $M\equiv 0$ (mod 3), $P_a=0, P_b=2, m_p\equiv 1, 0$
(mod 3) or}}
\leftline{\hskip12pt\underbar{$M\equiv 2$(mod 3), $P_a=2, P_b=0, m_p\equiv0, 2$
(mod 3)}}
\vskip .5cm
$$\eqalign{N(m_+,m_-,m_{2s})&=
{{\Bigl[{{M-m_p}\over 3}\Bigr] +m_-}\choose{m_+}}
{{\Bigl[{{2M+m_p-2}\over 3}\Bigr]-m_+-m_{2s}}\choose{m_-}}\times \cr
&\hbox{\hskip30pt}
{{\Bigl[{{2M+m_p-2}\over 3}\Bigr]-m_+-m_--m_{2s}}\choose{m_{2s}}}}\eqno(3.20)
$$
\vskip 1cm
\leftline{\underbar{c. $M\equiv 0$ (mod 3), $P_a=1, P_b=0,m_p\equiv0,2$
(mod 3) or}}
\leftline{\hskip12pt\underbar{
$M\equiv 2$ (mod 3), $P_a=0, P_b=1, m_p\equiv 1,2$ (mod3)}}
\vskip .5cm
$$\eqalign{N(m_+,m_-,m_{2s})&=
{{\Bigl[{{M-m_p+2}\over 3}\Bigr] +m_-}\choose{m_+}}
{{\Bigl[{{2M+m_p-2}\over 3}\Bigr]-m_+-m_{2s}}\choose{m_-}}\times \cr
&\hbox{\hskip30pt}{{\Bigl[{{2M+m_p-2}\over
3}\Bigr]-m_+-m_--m_{2s}}\choose{m_{2s}}}}\eqno(3.21)
$$
In all cases the exclusion rule (3.15) holds.

In the appendix we show for arbitrary $M$ and $Q$ that
$$\sum_{m_+,m_-,m_{2s}} 2^{m_{E}} N(m_+,m_-,m_{2s}) = 3^{M-1}. \eqno(3.22) $$
Thus these rules correctly count the number of states for all M.

\vfill
\eject
\font\bigmath=cmmib10 at 16pt

\textfont1=\bigmath
\leftline{\bgbf 4. Completeness for the
 $S=$1,$ \ \gamma =\pi /$3, XXZ spin chain}
\textfont1=\teni
\vskip .5cm
In section 2 we remarked that the Bethe's equation (2.29)  of the 3 state
superintegrable  chiral Potts model  is (up to a possible phase factor)
identical with   the Bethe's equation   which arises in the solution of
the $S=1$ anisotropic $XXZ$ chain with $\gamma = \pi/3$. Thus it is to
be expected that there is a close connection between these two models.

In the superintegrable chiral Potts model   the eigenvalues  are grouped
into sets  of size $2^{m_E}$. Each eigenvalue of the set is specified by
the same solution  of the Bethe's equation (2.29) and they differ  by the
$2^{m_E}$  independent choices      of the $m_E$ $ \pm$ signs in
(2.21)    and (2.25). We   will see in this section    that a completely
analagous phenomenon occurs  in the $XXZ$ chain with $\gamma = \pi /3$.
Namely, to each solution   of   the Bethe's equation (2.29) there also
corresponds   a set of eigenvalues   for   the $XXZ$ hamiltonian. However,
in distinction  with the chiral Potts chain, the eigenvalues in these
sets for    the $XXZ$ chain are degenerate.

The $S=1$ anisotropic integrable spin chain is defined by the hamiltonian [9]
$$\eqalign{H&=-\sum_{j=1}^{M}\Bigl\{ {\vec{S}}_j \cdot {\vec{S}}_{j+1} -
({\vec{S}}_j\cdot
 {\vec{S}}_{j+1})^2\cr
&\hbox{\hskip 30pt} - 2 \sin^2
{\gamma}[S_j^z  S_{j+1}^z - (S_j^z S_{j+1}^z)^2+ 2 (S_j^z)^2] \cr
&\hbox{\hskip30pt} - 2
(\cos\gamma-1)[(S_j^x  S_{j+1}^x+S_j^y  S_{j+1}^y)S_j^z S_{j+1}^z\cr
&\hbox{\hskip30pt} + S_{j}^{z} S_{j+1}^{z}(S_j^x  S_{j+1}^x+S_j^y
S_{j+1}^y)]\Bigr\}}
\eqno(4.1)$$
where periodic boundary conditions are imposed and the single site $3\times 3$
matrices $S^{x,y,z}$ are
$$\eqalignno{S^x &= {1 \over \sqrt 2}\pmatrix{0&1&0\cr 1&0&1 \cr
0&1&0},&(4.2a)\cr
S^y &= {i \over \sqrt 2}\pmatrix{0&-1&0\cr 1&0&-1 \cr 0&1&0\cr},&(4.2b)\cr
\noalign{\hbox{and}}
S^z &= \pmatrix{1&0&0 \cr 0&0&0 \cr 0&0&-1\cr}.&(4.2c)\cr}$$
The operator
$$S^z=\sum_{j=1}^{M} S_j^z\eqno(4.3)$$
commutes with (4.1) because of rotational invariance in the XY plane. The
eigenvalues of this spin chain have been extensively studied [12-15]
and in particular it is shown in refs. [12] and [13] that the energy
eigenvalues are
given by
$$E=\sin^2{(2\gamma)}\sum_{j=1}^{M-|S^z|}{ 1 \over \sinh{{1\over2}(\lambda_j
+ 2 i \gamma)} \sinh{{1\over2}(\lambda_j-2 i \gamma)}}
\eqno(4.4)$$
and the corresponding momentum eigenvalue is given by

$$e^{i P}= \prod_{j=1}^{M-|S^z|} {\sinh{{1\over2}(\lambda_j + 2 i \gamma)}
\over
\sinh{{1\over2}(\lambda_j - 2 i \gamma)}}
\eqno(4.5)$$
where the $\lambda_k$ satisfy the Bethe's equation
$$\Bigl[{\sinh{{1\over2}(\lambda_k+2 i \gamma)}\over\sinh{{1\over2}(\lambda_k-2
i \gamma)}}\Bigr]^M =
\prod_{l \neq k}^{M-|S^z|}{\sinh{{1\over2}(\lambda_k-\lambda_l+2 i \gamma)}
\over\sinh{{1\over2}(\lambda_k-\lambda_l-2 i \gamma)}}.
\eqno(4.6)$$
When $\gamma =\pi /3$ this Bethe's equation is identical with the equation
(2.29) of the superintegrable chiral Potts model for the cases where
$M-P_a-P_b-m_p\equiv 0$ (mod 3).

The spectrum and thermodynamics of (4.1) have been extensively studied
in refs. [12-15] and for $\gamma =0$ the question of completeness
has been solved in ref.[29, 30]. When $\gamma / \pi$ is rational a set
of degeneracies between energy levels occurs which
can be studied   by means of the quantum algebra $U_q[SU(2)]$ at a root
of unity [31,32]. However, for our purposes here of comparing the $XXZ$ chain
with the superintegrable chiral  Potts model we will not use this general
formalism  and will content ourselves with the examination of a six site chain.

We  consider  first those cases  where  the phase factor in equation (2.29)
is  unity. From table  1 we see that this occurs in $Q=0$ for $m_p=0,3,$and
$6$; in $Q=1$   for $P_a=2$, $P_b=0$, $m_p=1,4$ and for $P_a=0$, $P_b=1$,
$m_p=2,5$ and; in $Q=2$        for $P_a=0$, $P_b=2$, $m_p=1,4$ and for
$P_a=1$, $P_b=0$, $m_p=2,5$. Thus all  values of $m_p=M-|S^z|$ occur.

To compare the number of states in the $XXZ$ chain for given $S^z$ and $P$ with
the number of allowed solutions of the corresponding superintegrable
chiral Potts chain we must relate $m_p$ to $S^z$ and $P_{CP}$ to $P_{S=1}$.
The correspondence $m_p=M-|S^z|$ is valid for all $m_p$.
To find the correspondence
of momentum we compare the momentum formula of chiral Potts (2.27) with the
momentum formula of the S=1 chain (4.5) to obtain
$$e^{i P_{CP}}=e^{i P_{S=1}} \omega^{m_p-P_b}. \eqno(4.7)$$
Thus we find  for all cases where $M-P_a-P_b-m_p\equiv0$ (mod 3) that
$$
\hbox{if}\ m_p \equiv 0 \ (\hbox{mod}\ 3) \ \hbox{then} \
P_{S=1}=P_{CP}\eqno(4.8a)
$$
and
$$
\hbox{if}\ m_p \equiv 1, 2 \ (\hbox{mod}\ 3) \ \hbox{then} \ P_{S=1}=P_{CP}
-{2 \pi \over 3}.
\eqno(4.8b) $$
Using these correspondences we list in table 3 the number of states of the
$S=1$ model and the number of allowed solutions of (2.29) for the
superintegrable chiral Potts model obtained from table 1 of ref. [7]. In
terms of the $P_{S=1}$ defined in (4.8) the number of chiral Potts states
is symmetric in $\pm P_{S=1}$ even though for $m_p\equiv 1,2$ (mod3)
 they are not
symmetric in $P_{CP}$.

It is obvious from table 3 that the number of eigenvalues of the $S=1,\gamma
=\pi/3$ XXZ chain is significantly larger that the number of solutions of the
Bethe's equation (2.29) allowed by the counting rules of sec.3. The question
is thus to find where the extra solutions of the $XXZ$ chain come from.

In table 4 we  compute all eigenvalues of the $XXZ$ chain (4.1) with
$\gamma = \pi /3$ for a chain of six sites. We have also computed that subset
of eigenvalues obtained by substituting in the energy  formula  (4.4) the
solutions of (2.29) with phase factor one  which exist   for the
superintegrable chiral Potts chain. This subset of eigenvalues is underlined
in table 4.

For further discussion it  is necessary to consider $|S^z|=0,3$ and $6$
separately from $|S^z|=1, 2, 4,$ and $5$.

First consider $|S^z|=0,3,$ and $6$. Inspection  of table  4 shows that the
single underlined
eigenvalue  obtained  in $P=0$, $S^z=6$ from the superintegrable
chiral Potts model recurs   4  times in $S^z=3$, 6 times in $S^z=0$,
3 times in $S^z=-3$  and once in $S^z=-6$ for a total multiplicity of
$2^4$. Similarly the  underlined eigenvalues  which occur in $S^z=3$ recur
twice in $S^z=0$ and once in $S^z=-3$ for a total multiplicity of $2^2$.
This is an illustration   of the rule that an eigenvalue obtained from
the chiral Potts chain which occurs  singly in the sector $S^z_{max}$ will
recur in all sectors $S^z$ congruent to $S^z_{max}$ mod 3  with
a total multiplicity
$$2^{2S^z_{max}/3}=2^{2(M-m_p)/3}.\eqno(4.9)$$
This is precisely the factor $2^{m_E}$ which occurs in the superintegrable
completeness sum (3.22) and hence the computation of section 3 shows that
the number of levels in the sector $S^z\equiv 0 ($mod$3)$ is $3^{M-1}$.

 From the point of view of the Bethe's equation (4.6) these $2^{2S^z_{max}/3}$
degenerate eigenvalues correspond to solutions obtained by letting some
of the $\lambda_k$ go to $\pm \infty$. This mechanism of multiple occupancy
of $\lambda_k=\pm \infty$ is the same phenomenon that occurs in the
isotropic case $\gamma = 0$ for the cases $S=1$ [29,30] and $S=1/2$
[24,26,27] and in the Hubbard model [33].

For the case $S^z=1,2 ($mod$3)$ the situation is slightly more complicated.
Consider first $S^z=5$. Here table 4 tells us that in each of $P=0,\pm
2 \pi/6$ and $\pi$ there is a single level which comes from the superintegrable
chiral Potts model and that this level recurs twice in $S^z=2$ and once
in $S^z=-1$ (and similarly the states in $S^z=-5$ recur twice in
$S^z=-2$ and once in $S^z=1$).  Thus the total multiplicity is $2^2$.
Thus we have an identification of all the chiral Potts states in $Q=1$ and 2
and $m_p\equiv 1,2 ($mod$3)$ with the states in $S=1$.

We have now exhausted all the chiral Potts states with phase factor one.
However not all the states of the $S=1$ model have been been accounted for.
To obtain the remaining levels we must consider the remaining solutions
of the chiral Potts equation (2.29) where the phase factor is
$\omega ^{\pm 1}$ (see table 1). We see in table 4 that the eigenvalue
18 in $P=\pm 2 \pi/3$ for $S^z=5(-5)$ recurs 3 times in $S^z=2(-2)$, 3 times
in $S^z=-1(1)$  and once in $S^z=-4(4)$ for a total multiplicity of $2^3$.
This is identical with the multiplicity of the eigenvalues of table 2 with
$Q=1,P_a=2,P_b=0,m_p=0$ and $Q=2,P_a=0,P_b=2,m_p=0$. The remaining eigenvalues
are degenerate in $S^z=2(-1)$ and $-1(1)$ and are obtained by substituting
the chiral Potts values of $\lambda_k$ into the energy formula (4.4) . Thus
we have demonstrated for $M=6$ that all the eigenvalues of the $S=1$
chain with $\gamma = \pi/3$ can be obtained from the same solutions of the
Bethe's equation (2.29) that gave the superintegrable chiral Potts
eigenvalues. The degeneracies in the $S=1$ chain correspond to multiple
occupation of $\lambda_k = \pm \infty$ and the same exclusion rule (3.15)
found for chiral Potts holds also in the $XXZ$ chain. This phenomena holds
for all $M$. In general the $2^{m_E}$ states which correspond to the
choice of the $m_E$ $\pm$ signs in the chiral Potts model give in the
$S=1$ model degenerate eigenvalues for values of $S^z$ which are congruent
mod 3 and in each $S^z$ sector the degeneracy is given by a binomial
coefficient.

\vfill
\eject
\vskip 1 cm
\leftline{\bgbf 5. Excitations in the massive phase}
\vskip .5cm
In refs. [6,7] it was shown that the superintegrable chiral Potts chain (2.1)
has a massive phase if
$$ 0 \leq \lambda < \lambda_c = .9013...\ \hbox{and}\ 1 < \lambda
< \lambda_c^{-1}. \eqno(5.1)$$
In this phase, the spectrum of excitations for the sector
$Q=0$ is computed in detail.  In this section, we combine the computations of
ref. [7] with the completeness rules of section 3 to explore the physics
of the model.

We begin by recalling that if for $Q=0$ we consider the excitations in
(2.25) where all minus signs are chosen, then the result (4.9)
of ref. [7] for the energy eigenvalues is
$$\eqalign{\lim_{M\rightarrow\infty} \{E(P,\lambda,m_p)-E_{GS}\} &=
2m_p|1-\lambda|\cr
&+ {3 \over \pi} \int_1^{|{1+\lambda\over 1-\lambda}|^{2/3}}
dt \sum_{l=1}^{m_p}\Bigl\{{\omega v_l\over \omega t v_l -1} +
{\omega^2 v_l \over \omega^2 t v_l -1}\Bigr\}\times\cr
&\hbox{\hskip30pt}\Bigl[ {4 \lambda \over t^3-1}-(1-\lambda)^2\Bigr]^{1/2}}
\eqno(5.2)$$
and the corresponding momentum is
$$e^{-i P} = \prod_{l=1}^{m_p} \Bigl({1-\omega^2 v_l \over 1-\omega v_l}\Bigr)
\eqno(5.3) $$
(where we restrict our attention to $M\equiv0$ (mod 3) so that $m_p\equiv0$
(mod 3) and $P_a=P_b=0$).

To make contact with the completeness discussion of section 3, we consider
$v^+$ and $v^-$ together as
$$v=v^r \ \hbox{where} \ -\infty < v^r < \infty \eqno(5.4a) $$
and the 2-strings as
$$
v=v^{2s} e^{\pm \pi i/3} \ \hbox{where}\ v^{2s} \geq 0 \eqno(5.4b) $$
and define
$$\eqalign{
e_r &=2|1-\lambda| + {3 \over \pi} \int_1^{|{1+\lambda \over 1-\lambda}|^{2/3}}
dt \Bigl\{ {\omega v^r \over \omega t v^r -1} + {\omega^2 v^r \over
\omega^2 t v^r -1} \Bigl\} \times\cr
&\hbox{\hskip120pt} \Bigl[ {4 \lambda\over t^3 -1}  -
(1-\lambda)^2\Bigl]^{1/2}} \eqno (5.5a)
$$
and
$$\eqalign{
e_{2s} &= 4 |1-\lambda| + {3\over\pi}\int_1^{|{1+\lambda\over1-\lambda}|^{2/3}}
dt {v^{2s} [ 4 (v^{2s}t)^2 - v^{2s}t + 1]\over(v^{2s} t)^3 +1} \times\cr
&\hbox{\hskip120pt}\Bigl[{4 \lambda \over t^3-1}-(1-\lambda)^2\Bigr]^{1/2}.}
\eqno(5.5b)$$
Thus we rewrite (5.2) as
$$
\lim_{M\rightarrow\infty} \{ E(P,\lambda,m_r,m_{2s})-E_{GS}\}
=\sum_{l=1}^{m_r} e_r(v_l^r) + \sum_{l=1}^{m_{2s}} e_{2s}(v_l^{2s})\eqno(5.6)$$
where
$$m_r = m_+ + m_- \eqno(5.7) $$
and the $v_l^r$ and $v_l^{2s}$ satisfy the exclusion rules $v_l^r\neq v_k^r$
and $v_l^{2s} \neq v_k^{2s}$ .
Similarly, we rewrite (5.3) by defining
$$ e^{-i P^r} = \Bigl( {1-\omega^2 v^r \over 1-\omega v^r}\Bigr)\eqno(5.8a)$$
and
$$\eqalign{e^{-i P^{2s}} &=
\Bigl({1-\omega^2 e^{\pi i / 3} v^{2s}\over 1-\omega e^{\pi i/3}v^{2s}}\Bigr)
\Bigl({1-\omega^2 e^{-\pi i/ 3} v^{2s}\over 1-\omega e^{-\pi i/3}v^{2s}}\Bigr)
\cr &=\Bigl({1- e^{-\pi i / 3} v^{2s} \over 1- e^{\pi i/3}v^{2s}}\Bigr).}
\eqno(5.8b)$$
Thus (5.3) becomes
$$P=\sum_{l=1}^{m_r} P_l^r + \sum_{l=1}^{m_{2s}} P_l^{2s}\ (\hbox{mod}\
2\pi)\eqno(5.9)
$$
and we note from (5.4) and (5.8) that
$$ 0 \leq P^r < 2 \pi \eqno(5.10a)$$
whereas
$$
{2 \pi \over 3} \leq P^{2s} \leq 2\pi. \eqno(5.10b)$$
When (5.1) holds, both $e_r$ and $e_{2s}$ are positive.  The phase transition
 at $\lambda_c^{\pm 1}$ occurs because the gap in $e_r$ (but not in $e_{2s}$)
vanishes [6,7].

To complete the presentation of the energy spectrum of (2.1) it
remains to consider replacing an arbitrary number of minus signs in
(2.25) by plus signs. This is equivalent to adding $2w_l$ to
the energy where $w_l$ is given by (2.23) and $t_l$ are the roots of
the polynomial (2.24) in the ground state $m_p=0$. In
ref. [7] it is shown that in the ground state as $M\rightarrow \infty$
the roots are such that $t_l^3$ is less than minus one. Thus
we set $-t_l^3=(v^c_l)^3$ and define
$$e_c(v^c_l)=6\Big[{4\lambda\over1+(v^c_l)^3}+
(1-\lambda)^2\Bigr]^{1/2}\eqno(5.11)$$
with
$$1\leq v^c_l < \infty\eqno(5.12)$$
where we note that $e_c$ is positive for $\lambda\neq 1$.
Then, denoting the number of changed minus signs as $m_c$, we obtain the
complete explicit formula for the energy and momentum of all excitations
as
$$\lim_{M\rightarrow\infty}\{E_{ex}(P)-E_{GS}\} =\sum_{l=1}^{m_r}e_r(v_l^r)
+ \sum_{l=1}^{m_{2s}}e_{2s}(v_l^{2s}) +\sum_{l=1}^{m_c}e_c(v_l^c)\eqno(5.13a)$$
and
$$\lim_{M\rightarrow\infty} P=\sum_{l=1}^{m_r}P_l^r+\sum_{l=1}^{m_{2s}}P_l^{2s}
\eqno(5.13b)$$
where
$$-\infty < v^r < \infty, \ \ 0 < v^{2s},\ \ 1 < v^c ,\eqno(5.13c)$$
$$m_r+2m_{2s}\equiv0({\rm mod}3),\eqno(5.13d)$$
$e_r,\ e_{2s},\ e_c,\ P^r $ and $P^{2s}$ are given by (5.5), (5.11)
and (5.8), and the $v_l$ satisfy the exclusion rules $v_l^r\neq v_k^r$,
$v_l^{2s}\neq v_k^{2s}$ and $v_l^c\neq v_k^c$.
  In addition, we note that corresponding to the exclusion
rule $I_k^- \neq - I_l^{2s}$ (3.15), there will be an exclusion rule between
negative
$v^r$ and $v^{2s}$ of
$${-{1\over v_k^{r}}} \neq v_l^{2s}.\eqno(5.14)$$

We are now in a position to discuss the physics of the excitation
spectrum of the superintegrable chiral Potts model.  In particular,
we return to the discussion of the introduction and compare the
excitation spectrum (5.13) with the quasi-particle spectrum (1.5).

Consider first the special case $m_c=0$.  Then (5.13) has precisely
the form (1.5) where the excitation energies $e_r(P)$ and $e_{2s}(P)$
are obtained by eliminating $v^r$ and $v^{2s}$ between (5.5) and
(5.8).  Thus, using the definition of ``quasi-particle'' given in the
introduction, we may say that $e_r(P)$ and $e_{2s}(P)$ are the
energies of  single quasi-particle excitations
{}.

However, there are two striking differences between these
quasi-particles and all the previously discussed quasi-particle
excitations; namely, the momentum of the $2s$ excitations is only
allowed from (5.10b) to be in the range $2\pi/3\leq P^{2s} \leq 2\pi$ and from
(5.14) there is an exclusion rule between real
excitations of momentum $0\leq P^r\leq 4\pi/3$ and $2s$ excitations.  It thus
seems
completely fair to say that even though real and 2 string excitations
have the quasi-particle form (1.5) for energies and momentum, the
rules of combination are qualitatively different from those derived
from the intuition that is typically implied by the word
``quasi-particles.''

But even more striking than these new combination rules for energy
levels is the form that the energy levels take when $m_c\neq0$.  In
this case, we have a contribution $e_c(v^c)$ to the energy which
depends on one parameter but makes no contribution whatsoever to the
momentum.  This violates the quasi-particle form (1.5) and is what we
mean by non quasi-particle excitations.
\vfill
\eject

\vskip 1cm
\noindent{\bgbf 6. Non Quasi-Particle excitations and BCS Theory}
\vskip .5cm
The superintegrable chiral Potts model is the first system in which
non quasi-particle excitations, as defined in section 5, have been
observed in either exact or approximate computations. Nevertheless these
excitations have a striking resemblance to the Cooper pair excitations
of the BCS theory of superconductivity [16].

In section 3 of their famous 1957 paper [16] Bardeen, Cooper, and Schrieffer
classify the states of a superconducting system into three types:
single particle, ground pairs, and excited pairs. The implication of this
classification scheme is that single particle states and Cooper pairs are
logically distinct.

Single particle states are specified by a momentum ${\vec k}$. Excited
Cooper pairs are specified by a vector ${\vec k'}$. However the
momentum of the Cooper pair is zero. It is a most important result of
computation that in equation (2.52) of ref. [16] ``the energy to form
an excited pair in state ${\vec k'}$'' is $2E_{\vec k'}$ where
$E_{\vec k'}$ is the energy of a single particle excitation given by
equation (2.50) of ref [16]. The result of this computation is mandatory for
the interpretation of a Cooper pair as two quasi-particles with opposite
momenta.  This is the interpretation which allows the BCS
spectrum to be interpreted in the quasi-particle form (1.5).

But the clear reading of ref. [16] is that this identification of the energy
of the excited Cooper pair with $2E_{\vec k'} $is a result of
computation, and does not follow from model independent general principles.
The similar situation
would arise in the superintegrable chiral Potts model if
$$2e_r(P)=e_c(P).\eqno(6.1)$$
We have seen in section 5 that this
relation does not hold for the 3 state superintegrable chiral Potts
model. However, if we consider instead the two state case where the
chiral Potts model reduces to the Ising model the relation (6.1) does
indeed hold and, as is well known [3, 34, 35], the Ising model spectrum
does indeed have the quasi-particle form (1.5). Indeed, it is
significant that in ref. [35] Schulz, Mattis and Lieb note the strong
similarity of the Ising model with the classic works on
superconductivity [16, 36-38].

The results of this paper, of course, stand by themselves   without reference
to BCS theory. Nevertheless the above discussion of the presentation of
reference [16] suggests that the excitations $e_c$ can be thought of as the
generalization of Cooper pairs to a situation where the equality (6.1) fails.

\vfill
\eject
\vskip 1cm
\noindent{\bgbf 7. Conclusion}
\vskip .5cm
In this paper we have discussed three novel properties of the
excitation spectrum of the superintegrable chiral Potts model: 1)
single particle energies (5.5b) which do not exist for all ranges of
momentum (5.10b), 2) a fermion like exclusion rule (5.14) which
operates between different single quasi-particle levels and 3) non
quasi-particle excitations (5.11) which carry energy but make no contribution
to the momentum (5.13) and cannot be rewritten as the sum of single particle
levels.  Related phenomena certainly exist in the general chiral Potts
model (2.1) for $\phi \neq 0$. In addition the phenomenon of non
quasi-particle excitations continues to exist in the massless phase of
the chiral Potts model. Indeed, the phase transition at the point
$\lambda = 1$ [7] occurs because the gap in the non quasi-particle
energy $e_c$ vanishes.  Thus the chiral Potts model can be thought of
a providing a new universality class with novel physical effects which
have been seen first in theoretical studies before having been
experimentally observed.

On the other hand there may be reasons which forbid these effects
from ever appearing in real three dimensional systems. If this
is the case we hope that the present work will inspire an elucidation
of such a non-existence theorem.

\vfill
\eject
\vskip 1cm
\leftline{\bgbf Acknowledgements}
\vskip .5cm
We are pleased to acknowledge continuing discussions with G.
Albertini, H. Au-Yang, and J. H. H. Perk. We are also pleased to
acknowledge many useful discussions with P. Allen, R. J. Baxter, V. V.
Bazhanov, U. Grimm, J. Jain, P. A. Pearce and C. N. Yang.
  One of us (BMM) wishes to thank
R. J. Baxter and V. V. Bazhanov  for hospitality extended during a recent
visit to the Australian National University where part of this work
was done and to thank P. A. Pearce for hospitality extended during a
recent visit to the Univesity of Melbourne where this work
was completed. This work was partially supported by the National Science
Foundation under grant DMR-9106648.
\vfill\eject

\font\bgbf=cmbx10 at 14pt

\leftline{\bgbf References}
\vskip .5cm
\item{[1]}S.  Howes, L. P. Kadanoff and M. den Nijs, Nucl.
 Phys. B215 [FS7] (1983), 169.

\line{\hfill}
\item{[2]} G.  von Gehlen and V. Rittenberg, Nucl Phys. B257[FS14]
(1985), 351.

\line{\hfill}
\item{[3]}L. Onsager, Phys. Rev.65 (1944), 117.

\line{\hfill}
\item{[4]}G. Albertini, B. M. McCoy, J. H. H. Perk and S.Tang, Nucl.
 Phys. B314 (1989), 741.

\line{\hfill}
\item{[5]}R. J. Baxter, Physics Letts. A133 (1988), 185.

\line{\hfill}
\item{[6]} G. Albertini, B. M. McCoy and J. H. H. Perk, Phys Letts. A135
(1989), 159.

\line{\hfill}
\item{[7]} G. Albertini, B. M. McCoy and J. H. H. Perk in
Advanced Studies in Pure Mathematics vol. 19 (Kinokuniya-Academic Press
1989) 1.

\line{\hfill}
\item{[8]} B.  Davies, J. Phys. A23(1990), 2245; J. Math. Phys. 32 (1991),
2945 ; S. S. Roan Preprint Bonn MPI/91-70.

\line{\hfill}
\item{[9]} A. B. Zamolodchikov and V. A. Fateev, Sov. J. Nucl. Phys.
32 (1980), 298.

\line{\hfill}
\item{[10]}P. P. Kulish, N. Yu. Reshetikhin and E. K. Sklyanin,
Lett. Math. Phys. 5 (1981), 393.

\line{\hfill}
\item{[11]}P. P. Kulish and N. Yu. Reshetikhin, J. Sov. Math. 23 (1983)
, 2435.

\line{\hfill}
\item{[12]}K. Sogo, Phys. Letts. A104 (1984), 51.

\line{\hfill}
\item{[13]}A. N. Kirillov and N. Yu. Reshetikhin, J. Phys A20 (1987),
1565, 1587.

\line{\hfill}
\item{[14]}F.C. Alcaraz and M. J. Martins, J. Phys. A 22 (1989), 1829.

\line{\hfill}
\item{[15]}H. Frahm, N. C. Yu and M. Fowler, Nucl. Phys. B336
(1990), 396.

\line{\hfill}
\item{[16]}J. Bardeen, L. N. Cooper and J. R. Schrieffer,
Phys.Rev.108 (1957), 1175.

\line{\hfill}
\item{[17]}H. Au-Yang, B. M. McCoy, J. H. H. Perk, S. Tang and M. L.
Yan, Phys. Letts. A123 (1987), 219.

\line{\hfill}
\item{[18]}B. M. McCoy, J. H. H. Perk, S. Tang and C. H. Sah, Phys.
Letts. A125 (1987), 9.

\line{\hfill}
\item{[19]}H. Au-Yang, B. M. McCoy, J. H. H. Perk and S. Tang in vol. 1
of {\it Algebraic Analysis} (eds.) M. Kashiwaa and T. Kawai (1988), 29
(Academic
Press).

\line{\hfill}
\item{[20]}R. J. Baxter, J. H. H. Perk and H. Au-Yang, Phys. Letts.
A128 (1988), 138.

\line{\hfill}
\item{[21]}H. Au-Yang and J. H. H. Perk in Advanced Studies in Pure
Mathematics vol. 19 (Kinokuniya-Academic Press 1989), 57.

\line{\hfill}
\item{[22]}R. J. Baxter, V. V. Bazhanov and J. H. H. Perk,
Int. J. Mod. Phys. B 4 (1989), 803.

\line{\hfill}
\item{[23]}G. Albertini, S. Dasmahapatra, and B. M. McCoy, preprint
RIMS-834.

\line{\hfill}
\item{[24]}M. Takahashi, Prog. Theo. Phys. 46 (1971), 401.

\line{\hfill}
\item{[25]}M. Takahashi and M. Suzuki, Prog. Theo. Phys. 48 (1972), 2187.

\line{\hfill}
\item{[26]} H. Bethe, Z. f. Phys. 71 (1931), 205.

\line{\hfill}
\item{[27]}  L. D. Faddeev and L. Takhtajan, J. Sov. Math. 24 (1984), 241.

\line{\hfill}
\item{[28]} C. N. Yang and C. P. Yang, Phys. Rev. 150 (1966), 321.

\line{\hfill}
\item{[29]}A. N. Kirillov, J. Sov. Math. 30 (1985), 2298; 36 (1987),
115.

\line{\hfill}
\item{[30]}L. A. Takhtajan, Phys. Lett. A 87 (1982), 479.

\line{\hfill}
\item{[31]}D. Baranowski, V. Rittenberg, and G. Schutz, Nucl. Phys. B370
(1992), 551.

\line{\hfill}
\item{[32]}V. Pasquier and H. Saleur, Nucl. Phys. B330 (1990), 523.

\line{\hfill}
\item{[33]}F. H. L. Essler, V. E. Korepin and K. Schoutens, preprint ITP-SB
91-51.

\line{\hfill}
\item{[34]}B. Kaufman, Phys. Rev. 76 (1949) 1232.

\line{\hfill}
\item{[35]}T. D. Schulz, D. C. Mattis and E. H. Lieb, Rev. Mod. Phys. 36,
(1964),856.

\line{\hfill}
\item{[36]}N. N. Bogolubov, Nuovo Cimento 7 (1958) 794.

\line{\hfill}
\item{[37]}J. G. Valatin, Nuovo Cimento 7 (1958), 843.

\line{\hfill}
\item{[38]} P. W. Anderson, Phys. Rev. 112 (1958), 1900.

\vfill\eject

\font\sc=cmcsc10
\def\ss{\scriptstyle}
\font\bx=cmbx12 at 14pt

\line{\bx Appendix}
\line{\hfill}

We shall evaluate the sums (3.22), for all the possible sectors
detailed in section 3. Our method follows ref.[23, 24].

\line{\hfill}
\noindent$\underline{\bf Q=0}$

\line{\hfill}

\noindent Here the factor $N(m_{+}, m_{-}, m_{2s})$ of (3.22) is given
by (3.10).

\line{\hfill}

\noindent{\hskip -2pt $\bullet$} $\underline{M\equiv 0 (\hbox{mod 3}), P_a=0,
P_b=0, m_P\equiv 0 (\hbox{mod 3})}$.

\line{\hfill}

\noindent Set $M=3 l$, $m_P=3 p$. The corresponding value of $m_E$ is obtained
from table 1 to be $2(l-p)$. Also,
$$\Bigl[{2M+m_P \over3}\Bigr]=2l+p, \
\Bigl[{M+1-m_P\over3}\Bigr]=l-p,\eqno(A.1)$$
so that the sum (3.22) to be performed for this sector is written as
$$\eqalignno{S^{Q=0}_{M\equiv0}(l)&=\sum_{\ss p=0}^l \sum_{\ss m_{-}, m_{+},
m_{2s}} 2^{2(l-p)} {{2l+p-m_{+}-m_{2s}}\choose{m_{-}}}\times\cr&\qquad\qquad
{2l+p-m_{+}-m_{2s}-m_{-} \choose{m_{2s}}}\, {l-p+m_{-}\choose
m_{+}}\ \delta(2m_{2s}+m_{-}+m_{+}-3p),\cr&&(A.2a)\cr
&=\sum_{\ss p=0}^l 2^{2(l-p)}\Biggl\{\sum_{\ss m_{-}, m_{+},
m_{2s}}
{{2l-2p+m_{-}+m_{2s}}\choose{m_{-}}}\,{2l-2p+m_{2s}
\choose{m_{2s}}}\times\cr&\qquad\qquad\qquad
\ \times {l-p+m_{-}\choose m_{+}}\
\delta(2m_{2s}+m_{-}+m_{+}-3p)\Biggr\},&(A.2b)}$$
where $\delta(a)\equiv\delta_{a,0}$.
\noindent We use the relation
$${{2l-2p+m_{-}+m_{2s}}\choose{m_{-}}}\,{2l-2p+m_{2s}
\choose{m_{2s}}} = {{2l-2p+m_{-}+m_{2s}}\choose{m_{2s}}}\,{2l-2p+m_{-}
\choose{m_{-}}}\eqno(A.3)$$
to rewrite (A.2) as
$$\eqalignno{S^0_0(l)&=\sum_{p=0}^l 2^{2(l-p)} \ \sum_{\ss m_{-}} \
{2l-2p+m_{-} \choose{m_{-}}} \times&(A.4)\cr
&\qquad\qquad\times\sum_{\ss m_{+}, m_{2s}}
{{2l-2p+m_{-}+m_{2s}}\choose{m_{2s}}}\,{l-p+m_{-}\choose m_{+}}\
\delta(2m_{2s}+m_{-}+m_{+}-3p). }$$

\noindent Then the  following two identities,

$$\eqalignno{{1\over(1-x^2)^{2l-2p+m_{-}+1}}&=\sum_{\ss m_{2s}=0}^\infty
{{2l-2p+m_{-}+m_{2s}}\choose{m_{2s}}} \, x^{2m_{2s}}
&(A.5a)}$$
and
$$(1+x)^{l-p+m_{-}}=\sum_{m_{+}=0}^{l-p+m_{-}} {l-p+m_{-}\choose m_{+}} \,
x^{m_{+}}
\eqno(A.5b)$$

\noindent may be multiplied together, and using the Kroenecker
$\delta$ we get
$$S^0_0(l)=\sum_{p=0}^l 2^{2(l-p)} {1\over{2\pi i}}\oint {dx\over
x^{3p+1}} \,{1\over x^{(l-p)+1}}\, {1\over(1-x)^{2(l-p)+1}}\,\sum_{m_{-}}
{2l-2p+m_{-} \choose{m_{-}}}\, \Bigl({x\over
1-x}\Bigr)^{m_{-}}\eqno(A.6)$$
where the appropriate power of $x$ is picked out by the residue
theorem, with a contour which surrounds $x=0$ and no other
singularities.

\noindent Using
$${1\over(1-x)^A}\,\sum_{m_{-}=0}^{\infty}\,{A-1+m_{-}\choose m_{-}}\,
\Bigl({x\over 1-x}\Bigr)^{m_{-}}={1\over(1-2x)^A}, \eqno(A.7)$$
\noindent (A.6) reduces to
$$\eqalignno{S^0_0(l)&=\sum_{p=0}^l 2^{2(l-p)} {1\over 2\pi i} \oint
{dx\over x^{3p+1}} {1\over(1+x)^{(l-p)+1}}\,
{1\over(1-2x)^{2(l-p)+1}}\cr
&={1\over2\pi i}\oint dx {2^{2l}\over x(1+x)^{l+1} (1-2x)^{2l+1}}
\sum_{p=0}^l \Bigl( {(1+x)(1-2x)^2 \over 4x^3}\Bigr)^p,&(A.8)}$$
and the geometric sum is executed to give
$$S^0_0(l)={1\over2\pi i}\oint {dx\over x} \Bigl[
{(1-2x)\over(1-3x)x^{3l}} \ - \ {4^{l+1} x^3\over (1+x)^{l+1}
(1-2x)^{2l+1} (1-3x)}\Bigr].\eqno(A.9)$$
The second term in the square brackets does not have a single pole at
$x=0$, and hence does not contribute to the sum. The first term
then gives the desired answer
$$S^0_0(l)=3^{3l-1}=3^{M-1}. \eqno(A.10)$$

\noindent{\hskip -2pt $\bullet$} $\underline{M\equiv 1 (\hbox{mod 3}), P_a=2,
P_b=0, m_P\equiv 0 (\hbox{mod 3}) \hbox{or} P_a=0, P_b=1, m_P\equiv 2
(\hbox{mod 3})}$.

\line{\hfill}

\noindent Here we set $M=3l+1$. For $m_P\equiv0=3p$, $P_a=2$, $P_b=0$,
the number of states is
$$\eqalignno{S^0_{1, m_P\equiv0}(l)&=\sum_{\ss p=0}^l \sum_{\ss m_{-}, m_{+},
m_{2s}} 2^{2(l-p)}
{{2l+p-m_{+}-m_{2s}}\choose{m_{-}}}\times&(A.11a)\cr&\qquad\qquad
{2l+p-m_{+}-m_{2s}-m_{-} \choose{m_{2s}}}\, {l-p+m_{-}\choose
m_{+}}\ \delta(2m_{2s}+m_{-}+m_{+}-3p).}$$
There are no states for $m_P\equiv1$ and for $m_P\equiv2$, $P_a=0$,
$P_b=1$, the number of states is
$$\eqalignno{S^0_{1, m_P\equiv2}(l)&=\sum_{\ss p=1}^l \sum_{\ss m_{-}, m_{+},
m_{2s}} 2^{2(l-p)+1}
{{2l+p-m_{+}-m_{2s}}\choose{m_{-}}}\times&(A.11b)\cr&\qquad
{2l+p-m_{+}-m_{2s}-m_{-} \choose{m_{2s}}}\, {l-p+m_{-}+1\choose
m_{+}}\ \delta(2m_{2s}+m_{-}+m_{+}-3p+1),}$$
so that the total number of states is given by
$$S^0_1(l)=S^0_{1, m_P\equiv0}+S^0_{1,m_P\equiv2}.\eqno(A.11c)$$
Notice that (A.11a) is the same as (A.2). The sums in $S^0_{1,
m_P\equiv2}$ can be performed in the same way as $S^0_0(l)$, using
relations similar to (A.3), (A.5), (A.7), and the geometric sum in
(A.8), to give
$$S^0_{1,m_P\equiv2}=2 \, \cdot 3^{3l-1}.\eqno(A.12)$$
Therefore,
$$S^0_1(l)=3^{3l-1} + 2\cdot
3^{3l-1}\,=\,3^{3l}\,=\,3^{M-1}.\eqno(A.13)$$

\line{\hfill}
\noindent{\hskip -2pt $\bullet$} $\underline{M\equiv 2 (\hbox{mod 3}), P_a=1,
P_b=0, m_P\equiv 0 (\hbox{mod 3})\  \hbox{or}\  P_a=0, P_b=2, m_P\equiv 1
(\hbox{mod 3})}$.

\line{\hfill}

\noindent Set $M=3l+1$. For $m_P\equiv 0 =3p$, $P_a=1$, $P_b=0$,
$$\Bigl[{2M+m_P \over3}\Bigr]=2l+p+1, \
\Bigl[{M+1-m_P\over3}\Bigr]=l-p+1,\ m_E=2(l-p)+1 \eqno(A.14a)$$
while for $m_P\equiv 1=3p+1$, $P_a=0$, $P_b=2$,
$$\Bigl[{2M+m_P \over3}\Bigr]=2l+p+1, \
\Bigl[{M+1-m_P\over3}\Bigr]=l-p,\ m_E=2(l-p).\eqno(A.14b)$$
The total number of states is, therefore
$$\eqalignno{S^{Q=0}_{M\equiv2}(l)&=\sum_{\ss p=0}^l \Biggl\{\sum_{\ss
m_{-}, m_{+}, m_{2s}} 2^{2(l-p)+1}
{{2l-p+1-m_{2s}-m_{+}}\choose{m_{-}}}\,{l-p+1+m_{-}\choose m_{+}}
\times\cr&\qquad\qquad\qquad
\ \times {2l+p+1-m_{2s}-m_{+}-m_{-}
\choose{m_{2s}}}\
\delta(2m_{2s}+m_{-}+m_{+}-3p) \cr&\ +
\sum_{\ss
m_{-}, m_{+}, m_{2s}} 2^{2(l-p)}
{{2l+p+1-m_{+}-m_{2s}}\choose{m_{-}}}\,{2l+p+1-m_{2s}-m_{+}-m_{-}
\choose{m_{2s}}}\times\cr&\qquad\qquad\qquad
\ \times {l-p+m_{-}\choose m_{+}}\
\delta(2m_{2s}+m_{-}+m_{+}-3p-1)\Biggr\}.&(A.16)}$$
The sum is executed along the same lines as the steps (A.2) to (A.10)
performed to evaluate $S^0_0(l)$. The answer is
$$S^0_2(l)=3^{3l+1}=3^{M-1}.\eqno(A.17)$$

\line{\hfill}
\noindent$\underline{\bf Q=1}$

\line{\hfill}
\noindent{\hskip -2pt $\bullet$} $\underline{M\equiv 0 (\hbox{mod 3})}$

\line{\hfill}
Here, $N(m_{+}, m_{-}, m_{2s})$ is given by (3.18).
We set $M=3l$ and  $m_P=3p$,$3p+1$, $3p+2$ for $m_P\equiv0$, $1$ and
$2$ respectively. To evaluate the sum (3.22) for this case
there are three different terms to be combined.
$$\eqalignno{S^{Q=1}_{M\equiv0}(l)&=\sum_{\ss p=0}^{l-1} \Biggl\{\sum_{\ss
m_{-}, m_{+}, m_{2s}} 2^{2(l-p)}
{{2l+p-1-m_{2s}-m_{+}}\choose{m_{-}}}\,{l-p+m_{-}\choose m_{+}}
\times\cr&\qquad\qquad\qquad
\ \times {2l+p-1-m_{2s}-m_{+}-m_{-}
\choose{m_{2s}}}\
\delta(2m_{2s}+m_{-}+m_{+}-3p) \cr&\ +
\sum_{\ss
m_{-}, m_{+}, m_{2s}} 2^{2(l-p)-2}
{{2l+p-1-m_{+}-m_{2s}}\choose{m_{-}}}\,{l-p+m_{-}-1\choose
m_{+}}\times\cr&\qquad\qquad\qquad
\ \times \ {2l+p-1-m_{2s}-m_{+}-m_{-}
\choose{m_{2s}}}
\delta(2m_{2s}+m_{-}+m_{+}-3p-1)
\cr&\ +
\sum_{\ss
m_{-}, m_{+}, m_{2s}} 2^{2(l-p)-2}
{{2l+p-m_{+}-m_{2s}}\choose{m_{-}}}\,{l-p+m_{-}\choose m_{+}}
\times&(A.18)\cr&\qquad\qquad\qquad
\ \times \ {2l+p-m_{2s}-m_{+}-m_{-} \choose{m_{2s}}}
\delta(2m_{2s}+m_{-}+m_{+}-3p-2)\Biggr\}.}$$
We reorganise the binomial coefficients as in (A.3), to get
$$\eqalignno{S^{Q=1}_{M\equiv0}(l)&=\sum_{\ss p=0}^{l-1}
\Biggl\{2^{2(l-p)} \sum_{\ss m_{-}}
{{2l-2p-1+m_{-}}\choose{m_{-}}}\,  \times\cr&
\sum_{\ss m_{+}, m_{2s}}{l-p+m_{-}\choose m_{+}}\
{2l-2p-1+m_{2s}+m_{-}\choose{m_{2s}}}\
\delta(2m_{2s}+m_{-}+m_{+}-3p) \cr&\ \qquad + 2^{2(l-p)-2}
\sum_{\ss m_{-}}{{2l-2p-2+m_{-}}\choose{m_{-}}}\,\times&(A.19)\cr&
{l-p+m_{-}-1\choose m_{+}}\,{2l-2p-2+m_{2s}+m_{-}\choose{m_{2s}}}
\delta(2m_{2s}+m_{-}+m_{+}-3p-1)
\cr&\ \qquad + 2^{2(l-p)-2}
\sum_{\ss m_{-}}
{{2l-2p-2+m_{-}}\choose{m_{-}}}\,\times\cr&
\sum_{m_{+}, m_{2s}}{l-p+m_{-}\choose m_{+}}\,{2l+p-m_{2s}-m_{+}-m_{-}
\choose{m_{2s}}}
\delta(2m_{2s}+m_{-}+m_{+}-3p-2)\Biggr\},}$$
use the
generating functions (A.5) to perform the sum over $m_{+}$ and
$m_{2s}$ by picking out the appropriate residue, collapse the
Kroenecker $\delta$ and then use (A.7) to get
$$\eqalignno{\quad S^1_0(l)&=\quad{1\over 2\pi i}\oint
dx\,\Bigl[\sum_{p=0}^{l-1}
\Bigl({(1+x)(1-2x)^2\over 4x^3}\Bigr)^p\Bigr] \cdot \Bigl({4^l\over
(1+x)^l (1-2x)^{2l}}\Bigr)\cr&\ \times\Bigl\{{1\over x}+{(1-2x)\over
4x^2} + {(1+x)(1-2x)\over 4x^3}\Bigr\}\cr
&=\quad {1\over 2\pi i}\oint dx\,\Bigl( {1\over 3x^{3l}} - {4^l\over
(1+x)^l (1-2x)^{2l} }\Bigr)\cdot {1\over
(1-3x)}\cr&\qquad\qquad\times\Bigl\{4x^2+x(1-2x)+(1+x)(1-2x)\Bigr\} \cr
&=3^{3l-1}=3^{M-1}.&(A.20)}$$
\line{\hfill}
\noindent{\hskip -2pt $\bullet$} $\underline{M\equiv 1 (\hbox{mod 3})}$

\line{\hfill}
Here also, $N(m_{+}, m_{-}, m_{2s})$ is given by (3.18).
We set $M=3l+1$ and  $m_P=3p$,$3p-2$, $3p-1$ for $m_P\equiv0$, $1$ and
$2$ respectively. Thus we set up (3.22) as
$$\eqalignno{S^1_1(l)&=\sum_{\ss p=0}^{l} \sum_{\ss
m_{-}, m_{+}, m_{2s}} 2^{2(l-p)}
{{2l+p-m_{2s}-m_{+}}\choose{m_{-}}}\,{l-p+1+m_{-}\choose m_{+}}
\times&(A.21)\cr&\qquad\qquad\qquad
\ \times {2l+p-m_{2s}-m_{+}-m_{-}
\choose{m_{2s}}}\
\delta(2m_{2s}+m_{-}+m_{+}-3p) \cr&\ + \sum_{p=1}^l
\sum_{\ss
m_{-}, m_{+}, m_{2s}} 2^{2(l-p)+2}
{{2l+p-1-m_{+}-m_{2s}}\choose{m_{-}}}\,{l-p+m_{-}+1\choose
m_{+}}\times\cr&\qquad\qquad\qquad
\ \times \ {2l+p-1-m_{2s}-m_{+}-m_{-}
\choose{m_{2s}}}
\delta(2m_{2s}+m_{-}+m_{+}-3p+2)
\cr&\ + \sum_{p=1}^l
\sum_{\ss
m_{-}, m_{+}, m_{2s}} 2^{2(l-p)}
{{2l+p-1-m_{+}-m_{2s}}\choose{m_{-}}}\,{l-p+m_{-}\choose m_{+}}
\times\cr&\qquad\qquad\qquad
\ \times \ {2l+p-1-m_{2s}-m_{+}-m_{-} \choose{m_{2s}}}
\delta(2m_{2s}+m_{-}+m_{+}-3p+1).}$$
Going through the same steps as in the case $Q=1$, $M\equiv0$, we get
$$S^1_1(l)=3^{3l}.\eqno(A.22)$$

\line{\hfill}
\vfill\eject

\noindent{\hskip -2pt $\bullet$} $\underline{M\equiv 2 (\hbox{mod 3})}$

\line{\hfill}
Here, $N(m_{+}, m_{-}, m_{2s})$ is given by (3.17).
For this sector the sum to be evaluated is identical to the one in
$Q=0$, $M\equiv2$. Therefore, for $M=3l+2$,
$$S^1_2(l)=3^{3l+1}=3^{M-1}.\eqno(A.23)$$

\line{\hfill}
\noindent$\underline{\bf Q=2}$

\line{\hfill}
\noindent The sums are done in a fashion completely analogous to those
in $Q=1$.

\vfill\eject

\sm
\baselineskip=10pt
\centerline{{\bf Table 1.} The allowed values of $A+\lambda B$, $P_a$,
$P_b$, $m_E$, $m_P$, $3 m_E +2 m_P +P_a +P_b$, }
\centerline{ for the superintegrable 3-state chiral Potts model.}

\def\one{$\equiv 1~(\hbox{\rm mod}~3)$}
\def\two{$\equiv 2~(\hbox{\rm mod}~3)$}
\def\zero{$\equiv 0~(\hbox{\rm mod}~3)$}

\def\va{\vbox{{\kern.25em}
\hbox{$-3-\lambda$}
\hbox{$-4\lambda$}
\hbox{$+3-\lambda$}
\hbox{$+2\lambda$}}}

\def\vb{\vbox{{\kern.25em}
\hbox{$+3+\lambda$}
\hbox{$+4\lambda$}
\hbox{$-3+\lambda$}
\hbox{$-2\lambda$}}}


\def\ia{\vbox{{\kern.25em}
\hbox{$2$}
\hbox{$2$}
\hbox{$0$}
\hbox{$0$}}}

\def\ib{\vbox{{\kern.25em}
\hbox{$0$}
\hbox{$0$}
\hbox{$1$}
\hbox{$1$}}}

\def\ja{\vbox{{\kern.25em}
\hbox{$0$}
\hbox{$0$}
\hbox{$1$}
\hbox{$1$}}}

\def\jb{\vbox{
\hbox{$2$}
\hbox{$2$}
\hbox{$0$}
\hbox{$0$}}}

\def\mea{\vbox{{\kern.25em}
\hbox{$\equiv 1~(\hbox{\rm mod}~2)$}
\hbox{$\equiv 0~(\hbox{\rm mod}~2)$}
\hbox{$\equiv 1~(\hbox{\rm mod}~2)$}
\hbox{$\equiv 0~(\hbox{\rm mod}~2)$}}}

\def\meb{\vbox{{\kern.25em}
\hbox{$\equiv 1~(\hbox{\rm mod}~2)$}
\hbox{$\equiv 0~(\hbox{\rm mod}~2)$}
\hbox{$\equiv 1~(\hbox{\rm mod}~2)$}
\hbox{$\equiv 0~(\hbox{\rm mod}~2)$}}}

\def\mpa{\vbox{{\kern.25em}
\hbox{$\equiv 0~(\hbox{\rm mod}~3)$}
\hbox{$\equiv 1~(\hbox{\rm mod}~3)$}
\hbox{$\equiv 0~(\hbox{\rm mod}~3)$}
\hbox{$\equiv 2~(\hbox{\rm mod}~3)$}}}

\def\mpb{\vbox{{\kern.25em}
\hbox{$\equiv 0~(\hbox{\rm mod}~3)$}
\hbox{$\equiv 1~(\hbox{\rm mod}~3)$}
\hbox{$\equiv 0~(\hbox{\rm mod}~3)$}
\hbox{$\equiv 2~(\hbox{\rm mod}~3)$}}}

\def\sa{\vbox{{\kern.25em}
\hbox{$2M-1$}
\hbox{$2M-2$}
\hbox{$2M-2$}
\hbox{$2M-1$}}}

\def\sb{\vbox{{\kern.25em}
\hbox{$2M-1$}
\hbox{$2M-2$}
\hbox{$2M-2$}
\hbox{$2M-1$}}}

\def\pha{\vbox{{\kern.25em}
\hbox{\one}
\hbox{\zero}
\hbox{\two}
\hbox{\zero}}}

\def\phb{\vbox{{\kern.25em}
\hbox{\one}
\hbox{\zero}
\hbox{\two}
\hbox{\zero}}}

\def\ph{\vbox{{\kern.25em}
\hbox{\zero}}}

\setbox\strutbox=\hbox{\vrule height12pt depth5pt width0pt}
\def\strut{\relax\ifmmode\copy\strutbox\else\unhcopy\strutbox\fi}

$$\vbox{\tabskip=0pt \baselineskip=10pt
\def\tablerule{\noalign{\hrule}}

\halign
to425.82857pt{&#\vrule&\strut~~$#$~~&#\vrule&\strut$~#~$&#\vrule&\strut$~#~$&#\vrule&\strut$~#~$&#\vrule&\strut$~#~$&#\vrule&\strut$~#~$&#\vrule&\strut$~#~$&#\vrule&\strut$~#~$&#\vrule&\strut$~#~$&#\vrule&\strut$~#~$&#\vrule&\strut$~#~$&#\vrule&\strut$~#~$&#\vrule\cr\tablerule
&\multispan{15}\hfil\strut $M\equiv 0~(\hbox{\rm mod}~3)$\hfil&\cr\tablerule
&Q&&A+\lambda B&&P_a&&P_b&&m_E&&m_P&&3m_E +2 m_P +P_a +P_b&&M-P_a -P_b
-m_P&\cr\tablerule
&0&&0&&0&&0&&\equiv 0~(\hbox{\rm mod}~2)&&\ph&&\hfil 2
M\hfil&&\ph&\cr\tablerule
&1&&\va&&\ia&&\ja&&\mea&&\mpa&&\sa&&\pha&\cr\tablerule
&2&&\vb&&\ib&&\jb&&\meb&&\mpb&&\sb&&\phb&\cr\tablerule}}$$


\def\vaa{\vbox{{\kern.25em}
\hbox{$-2-2 \lambda$}
\hbox{$+1+\lambda$}}}

\def\vab{\vbox{{\kern.25em}
\hbox{$-2$}
\hbox{$+1-3\lambda$}
\hbox{$+4$}
\hbox{$+1+3\lambda$}}}

\def\vac{\vbox{{\kern.25em}
\hbox{$-2+2 \lambda$}
\hbox{$+1-\lambda$}}}

\def\iaa{\vbox{{\kern.25em}
\hbox{$2$}
\hbox{$0$}}}

\def\iab{\vbox{{\kern.25em}
\hbox{$0$}
\hbox{$0$}
\hbox{$2$}
\hbox{$2$}}}

\def\iac{\vbox{{\kern.25em}
\hbox{$0$}
\hbox{$0$}}}

\def\jaa{\vbox{{\kern.25em}
\hbox{$0$}
\hbox{$1$}}}

\def\jab{\vbox{{\kern.25em}
\hbox{$0$}
\hbox{$0$}
\hbox{$2$}
\hbox{$2$}}}

\def\jac{\vbox{{\kern.25em}
\hbox{$0$}
\hbox{$0$}}}

\def\meaa{\vbox{{\kern.25em}
\hbox{$\equiv 0~(\hbox{\rm mod}~2)$}
\hbox{$\equiv 1~(\hbox{\rm mod}~2)$}}}

\def\meab{\vbox{{\kern.25em}
\hbox{$\equiv 0~(\hbox{\rm mod}~2)$}
\hbox{$\equiv 1~(\hbox{\rm mod}~2)$}
\hbox{$\equiv 0~(\hbox{\rm mod}~2)$}
\hbox{$\equiv 1~(\hbox{\rm mod}~2)$}}}

\def\meac{\vbox{{\kern.25em}
\hbox{$\equiv 0~(\hbox{\rm mod}~2)$}
\hbox{$\equiv 1~(\hbox{\rm mod}~2)$}}}

\def\mpaa{\vbox{{\kern.25em}
\hbox{$\equiv 0~(\hbox{\rm mod}~3)$}
\hbox{$\equiv 2~(\hbox{\rm mod}~3)$}}}

\def\mpab{\vbox{{\kern.25em}
\hbox{$\equiv 0~(\hbox{\rm mod}~3)$}
\hbox{$\equiv 1~(\hbox{\rm mod}~3)$}
\hbox{$\equiv 2~(\hbox{\rm mod}~3)$}
\hbox{$\equiv 1~(\hbox{\rm mod}~3)$}}}

\def\mpac{\vbox{{\kern.25em}
\hbox{$\equiv 0~(\hbox{\rm mod}~3)$}
\hbox{$\equiv 2~(\hbox{\rm mod}~3)$}}}

\def\saa{\vbox{{\kern.25em}
\hbox{$2M$}
\hbox{$2M$}}}

\def\sab{\vbox{{\kern.25em}
\hbox{$2M-1$}
\hbox{$2M-2$}
\hbox{$2M-2$}
\hbox{$2M-1$}}}

\def\sac{\vbox{{\kern.25em}
\hbox{$2M-2$}
\hbox{$2M-1$}}}

\def\phaa{\vbox{{\kern.25em}
\hbox{\two}
\hbox{\one}}}

\def\phab{\vbox{{\kern.25em}
\hbox{\zero}
\hbox{\two}
\hbox{\zero}
\hbox{\one}}}

\def\phac{\vbox{{\kern.25em}
\hbox{\one}
\hbox{\two}}}

\setbox\strutbox=\hbox{\vrule height12pt depth5pt width0pt}
\def\strut{\relax\ifmmode\copy\strutbox\else\unhcopy\strutbox\fi}

$$\vbox{\tabskip=0pt \baselineskip=10pt
\def\tablerule{\noalign{\hrule}}

\halign
to428pt{&#\vrule&\strut~~$#$~~&#\vrule&\strut$~#~$&#\vrule&\strut$~#~$&#\vrule&\strut$~#~$&#\vrule&\strut$~#~$&#\vrule&\strut$~#~$&#\vrule&\strut$~#~$&#\vrule&\strut$~#~$&#\vrule&\strut$~#~$&#\vrule&\strut$~#~$&#\vrule&\strut$~#~$&#\vrule&\strut$~#~$&#\vrule\cr\tablerule
&\multispan{15}\hfil\strut $M\equiv 1~(\hbox{\rm mod}~3)$\hfil&\cr\tablerule
&Q&&A+\lambda B&&P_a&&P_b&&m_E&&m_P&&3m_E +2 m_P +P_a +P_b&&M-P_a -P_b
-m_P&\cr\tablerule

&0&&\vaa&&\iaa&&\jaa&&\meaa&&\mpaa&&\saa&&\phaa&\cr\tablerule
&1&&\vab&&\iab&&\jab&&\meab&&\mpab&&\sab&&\phab&\cr\tablerule
&2&&\vac&&\iac&&\jac&&\meac&&\mpac&&\sac&&\phac&\cr\tablerule}}$$


\def\vba{\vbox{{\kern.25em}
\hbox{$-1-1 \lambda$}
\hbox{$+2+2 \lambda$}}}

\def\vbb{\vbox{{\kern.25em}
\hbox{$-1+1 \lambda$}
\hbox{$+2-2 \lambda$}}}

\def\vbc{\vbox{{\kern.25em}
\hbox{$-4$}
\hbox{$+1-3\lambda$}
\hbox{$+2$}
\hbox{$-1+3\lambda$}}}

\def\iba{\vbox{{\kern.25em}
\hbox{$1$}
\hbox{$0$}}}

\def\ibb{\vbox{{\kern.25em}
\hbox{$0$}
\hbox{$0$}}}

\def\ibc{\vbox{{\kern.25em}
\hbox{$2$}
\hbox{$2$}
\hbox{$0$}
\hbox{$0$}}}

\def\jba{\vbox{{\kern.25em}
\hbox{$0$}
\hbox{$2$}}}

\def\jbb{\vbox{{\kern.25em}
\hbox{$0$}
\hbox{$0$}}}

\def\jbc{\vbox{{\kern.25em}
\hbox{$0$}
\hbox{$0$}
\hbox{$1$}
\hbox{$1$}}}

\def\meba{\vbox{{\kern.25em}
\hbox{$\equiv 1~(\hbox{\rm mod}~2)$}
\hbox{$\equiv 0~(\hbox{\rm mod}~2)$}}}

\def\mebb{\vbox{{\kern.25em}
\hbox{$\equiv 1~(\hbox{\rm mod}~2)$}
\hbox{$\equiv 0~(\hbox{\rm mod}~2)$}}}

\def\mebc{\vbox{{\kern.25em}
\hbox{$\equiv 0~(\hbox{\rm mod}~2)$}
\hbox{$\equiv 1~(\hbox{\rm mod}~2)$}
\hbox{$\equiv 0~(\hbox{\rm mod}~2)$}
\hbox{$\equiv 1~(\hbox{\rm mod}~2)$}}}

\def\mpba{\vbox{{\kern.25em}
\hbox{$\equiv 0~(\hbox{\rm mod}~3)$}
\hbox{$\equiv 1~(\hbox{\rm mod}~3)$}}}

\def\mpbb{\vbox{{\kern.25em}
\hbox{$\equiv 0~(\hbox{\rm mod}~3)$}
\hbox{$\equiv 1~(\hbox{\rm mod}~3)$}}}

\def\mpbc{\vbox{{\kern.25em}
\hbox{$\equiv 0~(\hbox{\rm mod}~3)$}
\hbox{$\equiv 2~(\hbox{\rm mod}~3)$}
\hbox{$\equiv 1~(\hbox{\rm mod}~3)$}
\hbox{$\equiv 2~(\hbox{\rm mod}~3)$}}}

\def\sba{\vbox{{\kern.25em}
\hbox{$2M$}
\hbox{$2M$}}}

\def\sbb{\vbox{{\kern.25em}
\hbox{$2M-1$}
\hbox{$2M-2$}}}

\def\sbc{\vbox{{\kern.25em}
\hbox{$2M-2$}
\hbox{$2M-1$}
\hbox{$2M-1$}
\hbox{$2M-2$}}}

\def\phba{\vbox{{\kern.25em}
\hbox{\one}
\hbox{\two}}}

\def\phbb{\vbox{{\kern.25em}
\hbox{\two}
\hbox{\one}}}

\def\phbc{\vbox{{\kern.25em}
\hbox{\zero}
\hbox{\one}
\hbox{\zero}
\hbox{\two}}}

\setbox\strutbox=\hbox{\vrule height12pt depth5pt width0pt}
\def\strut{\relax\ifmmode\copy\strutbox\else\unhcopy\strutbox\fi}

$$\vbox{\tabskip=0pt \baselineskip=10pt
\def\tablerule{\noalign{\hrule}}

\halign
to428pt{&#\vrule&\strut~~$#$~~&#\vrule&\strut$~#~$&#\vrule&\strut$~#~$&#\vrule&\strut$~#~$&#\vrule&\strut$~#~$&#\vrule&\strut$~#~$&#\vrule&\strut$~#~$&#\vrule&\strut$~#~$&#\vrule&\strut$~#~$&#\vrule&\strut$~#~$&#\vrule&\strut$~#~$&#\vrule&\strut$~#~$&#\vrule\cr\tablerule
&\multispan{15}\hfil\strut $M\equiv 2~(\hbox{\rm mod}~3)$\hfil&\cr\tablerule
&Q&&A+\lambda B&&P_a&&P_b&&m_E&&m_P&&3m_E +2 m_P +P_a +P_b&&M-P_a -P_b
-m_P&\cr\tablerule

&0&&\vba&&\iba&&\jba&&\meba&&\mpba&&\sba&&\phba&\cr\tablerule
&1&&\vbb&&\ibb&&\jbb&&\mebb&&\mpbb&&\sbb&&\phbb&\cr\tablerule
&2&&\vbc&&\ibc&&\jbc&&\mebc&&\mpbc&&\sbc&&\phbc&\cr\tablerule}}$$

\vfill\eject

\centerline{{\bf Table 2.} The number of allowed sets of $v_k$
(denoted by $N($sets$)$ ) which }
\centerline{solve (2.22) and the corresponding number of eigenvalues,
$N($sets$)
\times 2^{m_E},$}
\centerline{ for each of the allowed sets of $P_a$ and $P_b$ given in table 1.}

\def\paaa{\vbox{{\kern.25em}
\hbox{$~0~$}
\hbox{$~$}}}

\def\paab{\vbox{{\kern.25em}
\hbox{$~2~$}
\hbox{$~2~$}
\hbox{$~0~$}
\hbox{$~0~$}}}

\def\paac{\vbox{{\kern.25em}
\hbox{$~0~$}
\hbox{$~0~$}
\hbox{$~1~$}
\hbox{$~1~$}}}

\def\pbaa{\vbox{{\kern.25em}
\hbox{$~0~$}\hbox{$~$}
}}

\def\pbab{\vbox{{\kern.25em}
\hbox{$~0~$}
\hbox{$~0~$}
\hbox{$~1~$}
\hbox{$~1~$}}}

\def\pbac{\vbox{{\kern.25em}
\hbox{$~2~$}
\hbox{$~2~$}
\hbox{$~0~$}
\hbox{$~0~$}}}

\def\meaa{\vbox{{\kern.25em}
\hbox{$~2~$}
\hbox{$~0~$}}}

\def\meab{\vbox{{\kern.25em}
\hbox{$~1~$}
\hbox{$~0~$}
\hbox{$~1~$}
\hbox{$~0~$}}}

\def\meac{\vbox{{\kern.25em}
\hbox{$~1~$}
\hbox{$~0~$}
\hbox{$~1~$}
\hbox{$~0~$}}}

\def\mpaa{\vbox{{\kern.25em}
\hbox{$~0~$}
\hbox{$~3~$}}}

\def\mpab{\vbox{{\kern.25em}
\hbox{$~0~$}
\hbox{$~1~$}
\hbox{$~0~$}
\hbox{$~2~$}}}

\def\mpac{\vbox{{\kern.25em}
\hbox{$~0~$}
\hbox{$~1~$}
\hbox{$~0~$}
\hbox{$~2~$}}}

\def\naa{\vbox{{\kern.25em}
\hbox{$~1~$}
\hbox{$~5~$}}}

\def\nab{\vbox{{\kern.25em}
\hbox{$~1~$}
\hbox{$~1~$}
\hbox{$~1~$}
\hbox{$~4~$}}}

\def\nac{\vbox{{\kern.25em}
\hbox{$~1~$}
\hbox{$~1~$}
\hbox{$~1~$}
\hbox{$~4~$}}}

\def\Naa{\vbox{{\kern.25em}
\hbox{$~4~$}
\hbox{$~5~$}}}

\def\Nab{\vbox{{\kern.25em}
\hbox{$~2~$}
\hbox{$~1~$}
\hbox{$~2~$}
\hbox{$~4~$}}}

\def\Nac{\vbox{{\kern.25em}
\hbox{$~2~$}
\hbox{$~1~$}
\hbox{$~2~$}
\hbox{$~4~$}}}

\def\qaa{\vbox{{\kern.25em}
\hbox{$~0~$}
\hbox{$~$}}}

\def\qab{\vbox{{\kern.25em}
\hbox{$~1~$}
\hbox{$~$}
\hbox{$~$}
\hbox{$~$}}}

\def\qac{\vbox{{\kern.25em}
\hbox{$~2~$}
\hbox{$~$}
\hbox{$~$}
\hbox{$~$}}}

\setbox\strutbox=\hbox{\vrule height12pt depth5pt width0pt}
\def\strut{\relax\ifmmode\copy\strutbox\else\unhcopy\strutbox\fi}

$$\vbox{\tabskip=0pt \baselineskip=10pt
\def\tablerule{\noalign{\hrule}}

\halign
to298.00174pt{&#\vrule&\strut~~$#$~~&#\vrule&\strut$~#~$&#\vrule&\strut$~#~$&#\vrule&\strut$~#~$&#\vrule&\strut$~#~$&#\vrule&\strut$~#~$&#\vrule&\strut$~#~$&#\vrule&\strut$~#~$&#\vrule&\strut$~#~$&#\vrule\cr\tablerule
&\multispan{13}\hfil\strut $M~ = ~3$\hfil&\cr\tablerule
&Q&&P_a&&P_b&&m_E&&m_P&&N \hbox{(sets)}&&N \hbox{(eigenvalues)}=2^{m_E}\times
N\hbox{(sets)}&\cr\tablerule
&\qaa&&\paaa&&\pbaa&&\meaa&&\mpaa&&\naa&&\Naa&\cr\tablerule
&\multispan{11}\hfil total \hfil&&9&\cr\tablerule
&\qab&&\paab&&\pbab&&\meab&&\mpab&&\nab&&\Nab&\cr\tablerule
&\multispan{11}\hfil total \hfil&&9&\cr\tablerule
&\qac&&\paac&&\pbac&&\meac&&\mpac&&\nac&&\Nac&\cr\tablerule
&\multispan{11}\hfil total \hfil&&9&\cr\tablerule}}$$
\def\paba{\vbox{{\kern.25em}
\hbox{$~2~$}\hbox{$~$}
\hbox{$~0~$}
}}

\def\pabb{\vbox{{\kern.25em}
\hbox{$~1~$}\hbox{$~$}
\hbox{$~1~$}
\hbox{$~0~$}
\hbox{$~0~$}
}}

\def\pabc{\vbox{{\kern.25em}
\hbox{$~0~$}\hbox{$~$}
\hbox{$~0~$}}}

\def\pbba{\vbox{{\kern.25em}
\hbox{$~0~$}\hbox{$~$}
\hbox{$~1~$}
}}

\def\pbbb{\vbox{{\kern.25em}
\hbox{$~0~$}\hbox{$~$}
\hbox{$~0~$}
\hbox{$~2~$}
\hbox{$~2~$}
}}

\def\pbbc{\vbox{{\kern.25em}
\hbox{$~0~$}\hbox{$~$}
\hbox{$~0~$}
}}

\def\meba{\vbox{{\kern.25em}
\hbox{$~2~$}
\hbox{$~0~$}
\hbox{$~1~$}}}

\def\mebb{\vbox{{\kern.25em}
\hbox{$~2~$}
\hbox{$~0~$}
\hbox{$~1~$}
\hbox{$~0~$}
\hbox{$~1~$}}}

\def\mebc{\vbox{{\kern.25em}
\hbox{$~2~$}
\hbox{$~0~$}
\hbox{$~1~$}}}

\def\mpba{\vbox{{\kern.25em}
\hbox{$~0~$}
\hbox{$~3~$}
\hbox{$~2~$}}}

\def\mpbb{\vbox{{\kern.25em}
\hbox{$~0~$}
\hbox{$~3~$}
\hbox{$~1~$}
\hbox{$~2~$}
\hbox{$~1~$}}}

\def\mpbc{\vbox{{\kern.25em}
\hbox{$~0~$}
\hbox{$~3~$}
\hbox{$~2~$}}}

\def\nba{\vbox{{\kern.25em}
\hbox{$~1~$}
\hbox{$~5~$}
\hbox{$~9~$}}}

\def\nbb{\vbox{{\kern.25em}
\hbox{$~1~$}
\hbox{$~8~$}
\hbox{$~3~$}
\hbox{$~3~$}
\hbox{$~3~$}}}

\def\nbc{\vbox{{\kern.25em}
\hbox{$~1~$}
\hbox{$~5~$}
\hbox{$~9~$}}}

\def\Nba{\vbox{{\kern.25em}
\hbox{$~4~$}
\hbox{$~5~$}
\hbox{$18~$}}}

\def\Nbb{\vbox{{\kern.25em}
\hbox{$~4~$}
\hbox{$~8~$}
\hbox{$~6~$}
\hbox{$~3~$}
\hbox{$~6~$}}}

\def\Nbc{\vbox{{\kern.25em}
\hbox{$~4~$}
\hbox{$~5~$}
\hbox{$18~$}}}

\def\qba{\vbox{{\kern.25em}
\hbox{$~0~$}
\hbox{$~$}
\hbox{$~$}}}

\def\qbb{\vbox{{\kern.25em}
\hbox{$~1~$}
\hbox{$~$}
\hbox{$~$}
\hbox{$~$}
\hbox{$~$}}}

\def\qbc{\vbox{{\kern.25em}
\hbox{$~2~$}
\hbox{$~$}
\hbox{$~$}}}

\setbox\strutbox=\hbox{\vrule height12pt depth5pt width0pt}
\def\strut{\relax\ifmmode\copy\strutbox\else\unhcopy\strutbox\fi}

$$\vbox{\tabskip=0pt \baselineskip=10pt
\def\tablerule{\noalign{\hrule}}

\halign
to298.00174pt{&#\vrule&\strut~~$#$~~&#\vrule&\strut$~#~$&#\vrule&\strut$~#~$&#\vrule&\strut$~#~$&#\vrule&\strut$~#~$&#\vrule&\strut$~#~$&#\vrule&\strut$~#~$&#\vrule&\strut$~#~$&#\vrule&\strut$~#~$&#\vrule\cr\tablerule
&\multispan{13}\hfil\strut $M~ = ~4$\hfil&\cr\tablerule
&Q&&P_a&&P_b&&m_E&&m_P&&N \hbox{(sets)}&&N \hbox{(eigenvalues)}=2^{m_E}\times
N\hbox{(sets)}&\cr\tablerule
&\qba&&\paba&&\pbba&&\meba&&\mpba&&\nba&&\Nba&\cr\tablerule
&\multispan{11}\hfil total \hfil&&27&\cr\tablerule
&\qbb&&\pabb&&\pbbb&&\mebb&&\mpbb&&\nbb&&\Nbb&\cr\tablerule
&\multispan{11}\hfil total \hfil&&27&\cr\tablerule
&\qbc&&\pabc&&\pbbc&&\mebc&&\mpbc&&\nbc&&\Nbc&\cr\tablerule
&\multispan{11}\hfil total \hfil&&27&\cr\tablerule}}$$
\vfill \eject

\centerline{{\bf Table 2} (cont'd.)}
\def\paca{\vbox{{\kern.25em}
\hbox{$~1~$}\hbox{$~$}
\hbox{$~0~$}\hbox{$~$}}}

\def\pacb{\vbox{{\kern.25em}
\hbox{$~0~$}
\hbox{$~$}
\hbox{$~0~$}\hbox{$~$}
}}

\def\pacc{\vbox{{\kern.25em}
\hbox{$~2~$}
\hbox{$~$}
\hbox{$~2~$}
\hbox{$~0~$}
\hbox{$~$}
\hbox{$~0~$}}}

\def\pbca{\vbox{{\kern.25em}
\hbox{$~0~$}
\hbox{$~$}
\hbox{$~2~$}\hbox{$~$}
}}

\def\pbcb{\vbox{{\kern.25em}
\hbox{$~0~$}
\hbox{$~$}
\hbox{$~0~$}\hbox{$~$}
}}

\def\pbcc{\vbox{{\kern.25em}
\hbox{$~0~$}
\hbox{$~$}
\hbox{$~0~$}
\hbox{$~1~$}
\hbox{$~$}
\hbox{$~1~$}}}

\def\meca{\vbox{{\kern.25em}
\hbox{$~3~$}
\hbox{$~1~$}
\hbox{$~2~$}
\hbox{$~0~$}}}

\def\mecb{\vbox{{\kern.25em}
\hbox{$~3~$}
\hbox{$~1~$}
\hbox{$~2~$}
\hbox{$~0~$}}}

\def\mecc{\vbox{{\kern.25em}
\hbox{$~2~$}
\hbox{$~0~$}
\hbox{$~1~$}
\hbox{$~2~$}
\hbox{$~0~$}
\hbox{$~1~$}}}

\def\mpca{\vbox{{\kern.25em}
\hbox{$~0~$}
\hbox{$~3~$}
\hbox{$~1~$}
\hbox{$~4~$}}}

\def\mpcb{\vbox{{\kern.25em}
\hbox{$~0~$}
\hbox{$~3~$}
\hbox{$~1~$}
\hbox{$~4~$}}}

\def\mpcc{\vbox{{\kern.25em}
\hbox{$~0~$}
\hbox{$~3~$}
\hbox{$~2~$}
\hbox{$~1~$}
\hbox{$~4~$}
\hbox{$~2~$}}}

\def\nca{\vbox{{\kern.25em}
\hbox{$~1~$}
\hbox{$23~$}
\hbox{$~4~$}
\hbox{$11~$}}}

\def\ncb{\vbox{{\kern.25em}
\hbox{$~1~$}
\hbox{$23~$}
\hbox{$~4~$}
\hbox{$11~$}}}

\def\ncc{\vbox{{\kern.25em}
\hbox{$~1~$}
\hbox{$~5~$}
\hbox{$~9~$}
\hbox{$~5~$}
\hbox{$16~$}
\hbox{$~9~$}}}

\def\Nca{\vbox{{\kern.25em}
\hbox{$~8~$}
\hbox{$46~$}
\hbox{$16~$}
\hbox{$11~$}}}

\def\Ncb{\vbox{{\kern.25em}
\hbox{$~8~$}
\hbox{$46~$}
\hbox{$16~$}
\hbox{$11~$}}}

\def\Ncc{\vbox{{\kern.25em}
\hbox{$~4~$}
\hbox{$~5~$}
\hbox{$18~$}
\hbox{$20~$}
\hbox{$16~$}
\hbox{$18~$}}}

\def\qca{\vbox{{\kern.25em}
\hbox{$~0~$}
\hbox{$~$}
\hbox{$~$}
\hbox{$~$}}}

\def\qcb{\vbox{{\kern.25em}
\hbox{$~1~$}
\hbox{$~$}
\hbox{$~$}
\hbox{$~$}}}

\def\qcc{\vbox{{\kern.25em}
\hbox{$~2~$}
\hbox{$~$}
\hbox{$~$}
\hbox{$~$}
\hbox{$~$}
\hbox{$~$}}}

\setbox\strutbox=\hbox{\vrule height12pt depth5pt width0pt}
\def\strut{\relax\ifmmode\copy\strutbox\else\unhcopy\strutbox\fi}

$$\vbox{\tabskip=0pt \baselineskip=10pt
\def\tablerule{\noalign{\hrule}}

\halign
to298.00174pt{&#\vrule&\strut~~$#$~~&#\vrule&\strut$~#~$&#\vrule&\strut$~#~$&#\vrule&\strut$~#~$&#\vrule&\strut$~#~$&#\vrule&\strut$~#~$&#\vrule&\strut$~#~$&#\vrule&\strut$~#~$&#\vrule&\strut$~#~$&#\vrule\cr\tablerule
&\multispan{13}\hfil\strut $M~ = ~5$\hfil&\cr\tablerule
&Q&&P_a&&P_b&&m_E&&m_P&&N \hbox{(sets)}&&N \hbox{(eigenvalues)}=2^{m_E}\times
N\hbox{(sets)}&\cr\tablerule
&\qca&&\paca&&\pbca&&\meca&&\mpca&&\nca&&\Nca&\cr\tablerule
&\multispan{11}\hfil total \hfil&&81&\cr\tablerule
&\qcb&&\pacb&&\pbcb&&\mecb&&\mpcb&&\ncb&&\Ncb&\cr\tablerule
&\multispan{11}\hfil total \hfil&&81&\cr\tablerule
&\qcc&&\pacc&&\pbcc&&\mecc&&\mpcc&&\ncc&&\Ncc&\cr\tablerule
&\multispan{11}\hfil total \hfil&&81&\cr\tablerule}}$$
\def\pada{\vbox{{\kern.25em}
\hbox{$~0~$}\hbox{$~~$}
\hbox{$~~$}
}}

\def\padb{\vbox{{\kern.25em}
\hbox{$~2~$}
\hbox{$~~$}
\hbox{$~2~$}
\hbox{$~~$}
\hbox{$~0~$}
\hbox{$~~$}
\hbox{$~0~$}\hbox{$~~$}
}}

\def\padc{\vbox{{\kern.25em}
\hbox{$~0~$}
\hbox{$~~$}
\hbox{$~0~$}
\hbox{$~~$}
\hbox{$~1~$}
\hbox{$~~$}
\hbox{$~1~$}\hbox{$~~$}
}}

\def\pbda{\vbox{{\kern.25em}
\hbox{$~0~$}\hbox{$~~$}
\hbox{$~~$}
}}

\def\pbdb{\vbox{{\kern.25em}
\hbox{$~0~$}
\hbox{$~~$}
\hbox{$~0~$}
\hbox{$~~$}
\hbox{$~1~$}
\hbox{$~~$}
\hbox{$~1~$}\hbox{$~~$}
}}

\def\pbdc{\vbox{{\kern.25em}
\hbox{$~2~$}
\hbox{$~~$}
\hbox{$~2~$}
\hbox{$~~$}
\hbox{$~0~$}
\hbox{$~~$}
\hbox{$~0~$}\hbox{$~~$}
}}

\def\meda{\vbox{{\kern.25em}
\hbox{$~4~$}
\hbox{$~2~$}
\hbox{$~0~$}}}

\def\medb{\vbox{{\kern.25em}
\hbox{$~3~$}
\hbox{$~1~$}
\hbox{$~2~$}
\hbox{$~0~$}
\hbox{$~3~$}
\hbox{$~1~$}
\hbox{$~2~$}
\hbox{$~0~$}}}

\def\medc{\vbox{{\kern.25em}
\hbox{$~3~$}
\hbox{$~1~$}
\hbox{$~2~$}
\hbox{$~0~$}
\hbox{$~3~$}
\hbox{$~1~$}
\hbox{$~2~$}
\hbox{$~0~$}}}

\def\mpda{\vbox{{\kern.25em}
\hbox{$~0~$}
\hbox{$~3~$}
\hbox{$~6~$}}}

\def\mpdb{\vbox{{\kern.25em}
\hbox{$~0~$}
\hbox{$~3~$}
\hbox{$~1~$}
\hbox{$~4~$}
\hbox{$~0~$}
\hbox{$~3~$}
\hbox{$~2~$}
\hbox{$~5~$}}}

\def\mpdc{\vbox{{\kern.25em}
\hbox{$~0~$}
\hbox{$~3~$}
\hbox{$~1~$}
\hbox{$~4~$}
\hbox{$~0~$}
\hbox{$~3~$}
\hbox{$~2~$}
\hbox{$~5~$}}}

\def\nda{\vbox{{\kern.25em}
\hbox{$~1~$}
\hbox{$46~$}
\hbox{$43~$}}}

\def\ndb{\vbox{{\kern.25em}
\hbox{$~1~$}
\hbox{$23~$}
\hbox{$~4~$}
\hbox{$11~$}
\hbox{$~1~$}
\hbox{$23~$}
\hbox{$19~$}
\hbox{$32~$}}}

\def\ndc{\vbox{{\kern.25em}
\hbox{$~1~$}
\hbox{$23~$}
\hbox{$~4~$}
\hbox{$11~$}
\hbox{$~1~$}
\hbox{$23~$}
\hbox{$19~$}
\hbox{$32~$}}}

\def\Nda{\vbox{{\kern.25em}
\hbox{$16~$}
\hbox{$184$}
\hbox{$43~$}}}

\def\Ndb{\vbox{{\kern.25em}
\hbox{$~8~$}
\hbox{$46~$}
\hbox{$16~$}
\hbox{$11~$}
\hbox{$~8~$}
\hbox{$46~$}
\hbox{$76~$}
\hbox{$32~$}}}

\def\Ndc{\vbox{{\kern.25em}
\hbox{$~8~$}
\hbox{$46~$}
\hbox{$16~$}
\hbox{$11~$}
\hbox{$~8~$}
\hbox{$46~$}
\hbox{$76~$}
\hbox{$32~$}}}

\def\qda{\vbox{{\kern.25em}
\hbox{$~0~$}
\hbox{$~$}
\hbox{$~$}}}

\def\qdb{\vbox{{\kern.25em}
\hbox{$~1~$}
\hbox{$~$}
\hbox{$~$}
\hbox{$~$}
\hbox{$~$}
\hbox{$~$}
\hbox{$~$}
\hbox{$~$}}}

\def\qdc{\vbox{{\kern.25em}
\hbox{$~2~$}
\hbox{$~$}
\hbox{$~$}
\hbox{$~$}
\hbox{$~$}
\hbox{$~$}
\hbox{$~$}
\hbox{$~$}}}

\setbox\strutbox=\hbox{\vrule height12pt depth5pt width0pt}
\def\strut{\relax\ifmmode\copy\strutbox\else\unhcopy\strutbox\fi}

$$\vbox{\tabskip=0pt \baselineskip=10pt
\def\tablerule{\noalign{\hrule}}

\halign
to298.00174pt{&#\vrule&\strut~~$#$~~&#\vrule&\strut$~#~$&#\vrule&\strut$~#~$&#\vrule&\strut$~#~$&#\vrule&\strut$~#~$&#\vrule&\strut$~#~$&#\vrule&\strut$~#~$&#\vrule&\strut$~#~$&#\vrule&\strut$~#~$&#\vrule\cr\tablerule
&\multispan{13}\hfil\strut $M~ = ~6$\hfil&\cr\tablerule
&Q&&P_a&&P_b&&m_E&&m_P&&N \hbox{(sets)}&&N \hbox{(eigenvalues)}=2^{m_E}\times
N\hbox{(sets)}&\cr\tablerule
&\qda&&\pada&&\pbda&&\meda&&\mpda&&\nda&&\Nda&\cr\tablerule
&\multispan{11}\hfil total \hfil&&243&\cr\tablerule
&\qdb&&\padb&&\pbdb&&\medb&&\mpdb&&\ndb&&\Ndb&\cr\tablerule
&\multispan{11}\hfil total \hfil&&243&\cr\tablerule
&\qdc&&\padc&&\pbdc&&\medc&&\mpdc&&\ndc&&\Ndc&\cr\tablerule
&\multispan{11}\hfil total \hfil&&243&\cr\tablerule}}$$

\vfill\eject

{{\bf  Table 3.} Comparison for a six-site chain of the number of
states of the $S=1$, XXZ model with the number of allowed solutions of
the superintegrable chiral Potts equation. The correspondence of
momenta is that for $m_P\equiv 0 (\hbox{mod} 3)$, $P_{S=1}=P_{CP}$
and for $m_P\equiv 1$, $2 (\hbox{mod} 3)$, $P_{S=1}=P_{CP}-2\pi/3$.
}
\setbox\strutbox=\hbox{\vrule height12pt depth5pt width0pt}
\def\strut{\relax\ifmmode\copy\strutbox\else\unhcopy\strutbox\fi}

$$\vbox{\tabskip=0pt \baselineskip=10pt
\def\tablerule{\noalign{\hrule}}
\halign to
330.7934pt{&#\vrule&\strut$~#~$&#\vrule&\strut$~#~$&#\vrule&\strut$~#~$&#\vrule&\strut$~#~$&#\vrule&\strut$#$&#\vrule&\strut$~#~$&#\vrule&\strut$~#~$&#\vrule&\strut$~#~$&#\vrule&\strut$~#~$&#\vrule&\strut$#$&#\vrule&\strut$#$&#\vrule&\strut$~#~$&#\vrule&\strut$#$&#\vrule&\strut$~#~$&#\vrule&\strut$~#~$&#\vrule&\strut$~#~$&#\vrule&\strut$~#~$&#\vrule&\strut$~#~$&#\vrule\cr\tablerule
&\multispan{19}\hfil\hbox{chiral
Potts}\hfil&&\,&&\multispan{13}\hfil \hbox{S=1 XXZ}\hfil&\cr\tablerule
&\multispan{19}$\hfil\hskip50pt P_{S=1}\hfil$&&\,&&\multispan{13}$\hfill
P_{S=1}\hfill$&\cr\tablerule
&m_P&&Q&&P_a&&P_b&&\!&&~0~&&\pm{2\pi\over6}&&\pm{4\pi\over6}&&~\pi~&&~total~&&\,&&S^z&&\!&&0&&\pm{2\pi\over6}&&\pm{4\pi\over6}&&\pi&&total&\cr\tablerule

&0&&0&&0&&0&&\,&&~1&&0&&0&&0&&~1~&&\,&&6&&\,&&~1&&~0&&~0&&~0&&~~1&\cr\tablerule
&1&&1&&2&&0&&\,&&~1&&1&&0&&1&&~4~&&\,&&5&&\,&&~1&&~1&&~1&&~1&&~~6&\cr\tablerule
&2&&1&&0&&1&&\,&&~4&&3&&3&&3&&~19~&&\,&&4&&\,&&~4&&~3&&~4&&~3&&~21&\cr\tablerule
&3&&0&&0&&0&&\,&&~5&&8&&8&&9&&~46~&&\,&&3&&\,&&~9&&~8&&~8&&~9&&~50&\cr\tablerule
&4&&1&&2&&0&&\,&&~4&&0&&3&&1&&~11~&&\,&&2&&\,&&16&&14&&16&&14&&~90&\cr\tablerule
&5&&1&&0&&1&&\,&&~6&&5&&5&&6&&~32~&&\,&&1&&\,&&21&&21&&21&&21&&126&\cr\tablerule
&6&&0&&0&&0&&\,&&10&&6&&8&&5&&~43~&&\,&&0&&\,&&26&&21&&24&&23&&141&\cr\tablerule
}}$$

\vfill\eject

\def\u#1{\underline{#1}}
\font\sc=cmcsc10

{\bf Table 4.} The eigenvalues of the $S=1$, $\gamma=\pi/3$ XXZ
periodic spin chain of six sites. The eigenvalues for negative values of
$S^z$ are obtained using the symmetry of $S^z \rightarrow -S^z$.
 The underlined eigenvalues are
obtained by using the solutions of (2.29) allowed for the
superintegrable chiral Potts model in the energy formula (4.4).

\def\sixa{\vbox{{\kern.25em}
\hbox{$\u{8.29991623268163}, 10.50604079256506, 10.50604079256507,
\u{10.94098877100413},$}
\hbox{$\u{13.08651430707976}, 13.89008373582524, 13.89008373582526, \u{14},
\u{14},$}
\hbox{$\u{14.76257547310441}, \u{15.61946171799351}, 16, 16, 16, 16,$}
\hbox{$16.60387547160966, 16.60387547160969, 18, 18, 18, 18, 18, 18,$}
\hbox{$\u{19.19636205699761}, \u{21.74114623721635}, \u{24.35303520392256}$}}}

\def\sixb{\vbox{{\kern.25em}
\hbox{$\u{9.56483858138269}, \u{12}, 12.15498460202909, 12.1549846020291,$}
\hbox{$\u{12.76393202250021}, \u{14.69952946524302}, 14.82874774570659,
14.82874774570659,$}
\hbox{$15.1684824307572, 15.16848243075721, 15.60102359882231,
15.60102359882231,$}
\hbox{$16.39000960599962, 16.39000960599963, \u{17.23606797749979},
17.91683389294189,$}
\hbox{$17.91683389294189, 18.09653253799186, 18.09653253799186,
19.84338558575142,$}
\hbox{$19.84338558575144, \u{21.73563195337429}$}}}

\def\sixc{\vbox{{\kern.25em}
\hbox{$\u{10.53589838486224}, \u{11.22809490982445}, \u{12.35288805589831},
13.13859446481223,$}
\hbox{$13.13859446481224, 13.29978403077409, 13.29978403077409,
\u{14.24071978009154},$}
\hbox{$14.30123932226858, 14.30123932226858, 15.52206260527575,
15.52206260527575,$}
\hbox{$16.49825145060049, 16.49825145060049, \u{16.55312816507534},
16.68317951010723,$}
\hbox{$16.68317951010724, \u{17.46410161513774}, \u{18.21877692510021},
\u{19.40639216401014},$}
\hbox{$20.07618217185877, 20.07618217185878, 20.48070644430281,
20.48070644430282$}}}

\def\sixd{\vbox{{\kern.25em}
\hbox{$\u{10.87689437438233}, 12.6791155005525, 12.67911550055251,
12.67911550055251,$}
\hbox{$12.67911550055252, 13.60074370344935, 13.60074370344937, \u{14},$}
\hbox{$14.64806079844652, 14.64806079844654, 14.64806079844654,
14.64806079844655,$}
\hbox{$\u{16}, \u{16}, 17.4782465565131, 17.47824655651312,
18.67282370100094,$}
\hbox{$18.67282370100094, 18.67282370100095, 18.67282370100095,$}
\hbox{$\u{19.12310562561766}, 21.92100974003751, 21.92100974003753$}}}

\def\fivea{\vbox{{\kern.25em}
\hbox{$\u{8.5214914003045}, \u{11.35042411530895}, 12.43669309493757,
12.43669309493761,$}
\hbox{$\u{13.37655732137895}, 14.26794919243112, 14.26794919243113, 15, 15,
15,$}
\hbox{$15.42532835575839, 15.42532835575842, \u{17}, \u{17},
17.73205080756889,$}
\hbox{$17.73205080756892, \u{18.75152716300766}, 19.10481891440092,$}
\hbox{$19.10481891440093, 21.03315963490303, 21.03315963490304$}}}

\def\fiveb{\vbox{{\kern.25em}
\hbox{$\u{9.8720197837459}, 10.76229969320372, \u{12.36017677219731},
13.12119617997308,$}
\hbox{$13.46167018036698, 13.94591578461394, 13.94591578461395,
14.82774302161003,$}
\hbox{$15.16604849636401, \u{15.52216410833847}, 16, \u{16.08917517655664},$}
\hbox{$16.78684626768899, 16.80912962588481, 17.11126415759022,
17.11126415759023,$}
\hbox{$18.8139813916125, \u{19.15646415916171}, 19.94282005779584,
19.94282005779586,$}
\hbox{$21.25108514329595$}}}

\def\fivec{\vbox{{\kern.25em}
\hbox{$\u{10.7817578014951}, 11.50761996219081, 12.20871215252207,
\u{12.77589846221546},$}
\hbox{$13.93039354102618, 14.15544528228914, 14.19806226419515,
14.19806226419517, $}
\hbox{$\u{14.77231645874024}, 15.55495813208737, 15.55495813208738,
16.79128784747792, $}
\hbox{$17.24697960371746, 17.24697960371748, \u{17.38267160768822}, 18, 18, $}
\hbox{$18, 19.02049822873401, \u{21.28735566986096}, 21.38604298575983$}}}

\def\fived{\vbox{{\kern.25em}
\hbox{$\u{11.13933320993551}, 12.55051025721681, 12.55051025721682, \u{13},
\u{13}$}
\hbox{$13.62771867673097, 13.62771867673099, \u{14.59566258403823}, $}
\hbox{$16, 16, 16, 16, 16, 16, 17.44948974278319, 17.44948974278321, $}
\hbox{$\u{18.51366656949581}, 19, 19.372281323269, 19.37228132326903, $}
\hbox{$\u{23.75133763653045}$}}}

\def\foura{\vbox{{\kern.25em}
\hbox{$\u{9.22646217887607}, 12.43669309493758, 14.26794919243112,
14.26794919243113, $}
\hbox{$\u{14.49943450055038}, 15, 15, 15, 15, 15.4253283557584,
\u{16.84789556123169}, $}
\hbox{$17.73205080756888, 17.7320508075689, 19.10481891440096, $}
\hbox{$21.03315963490305, \u{23.42620775934184}$}}}

\def\fourb{\vbox{{\kern.25em}
\hbox{$10.76229969320369, 13.12119617997307, 13.46167018036698,
13.94591578461395, $}
\hbox{$14.82774302161002, 15.166048496364, 16, 16, 16.78684626768898, $}
\hbox{$16.80912962588479, 17.11126415759021, 18.81398139161249, $}
\hbox{$19.94282005779584, 21.25108514329594$}}}

\def\fourc{\vbox{{\kern.25em}
\hbox{$11.50761996219082, \u{11.65982702674793}, 12.20871215252208,
13.93039354102619, $}
\hbox{$14.15544528228914, 14.19806226419516, \u{15.37778436506803},
15.55495813208736, $}
\hbox{$16.79128784747792, 17.24697960371747, 18, 18, 18,
\u{18.96238860818403},$}
\hbox{$19.02049822873401, 21.38604298575983$}}}

\def\fourd{\vbox{{\kern.25em}
\hbox{$\u{12}, 12.55051025721682, 12.55051025721683, 13.62771867673099, $}
\hbox{$13.62771867673099, 16, 16, 16, 17.44948974278317, 17.44948974278318,$}
\hbox{$19, 19, 19.37228132326903, 19.37228132326903$}}}

\def\threea{\vbox{{\kern.25em}
\hbox{$\u{10.50604079256506}, \u{13.89008373582526}, \u{16}, \u{16},
\u{16.60387547160968},$}
\hbox{$18, 18, 18, 18$}}}

\def\threeb{\vbox{{\kern.25em}
\hbox{$\u{12.15498460202909}, \u{14.82874774570659}, \u{15.16848243075719},
\u{15.6010235988223},$}
\hbox{$ \u{16.39000960599962}, \u{17.91683389294189}, \u{18.09653253799185},
\u{19.84338558575142}$}}}

\def\threec{\vbox{{\kern.25em}
\hbox{$\u{13.13859446481224}, \u{13.29978403077409}, \u{14.30123932226858},
\u{15.52206260527575},$}
\hbox{$\u{16.49825145060048}, \u{16.68317951010723}, \u{20.07618217185878},
\u{20.48070644430282}$}}}

\def\threed{\vbox{{\kern.25em}
\hbox{$\u{12.6791155005525}, \u{12.67911550055252}, \u{13.60074370344936},
\u{14.64806079844653},$}
\hbox{$\u{14.64806079844655}, \u{17.47824655651311}, \u{18.67282370100094},
\u{18.67282370100096},$}
\hbox{$\u{21.92100974003754}$}}}

\def\twoa{\vbox{{\kern.25em}
\hbox{$\u{12.4366930949376}, \u{15.42532835575841}, \u{19.10481891440097},
\u{21.03315963490304}$}}}

\def\twob{\vbox{{\kern.25em}
\hbox{$\u{13.94591578461395}, \u{17.11126415759022}, \u{19.94282005779584}$}}}

\def\twoc{\vbox{{\kern.25em}
\hbox{$\u{14.19806226419517}, \u{15.55495813208738}, \u{17.24697960371747},
18$}}}

\def\twod{\vbox{{\kern.25em}
\hbox{$\u{16}, \u{16}, \u{16}$}}}

\setbox\strutbox=\hbox{\vrule height12pt depth5pt width0pt}
\def\strut{\relax\ifmmode\copy\strutbox\else\unhcopy\strutbox\fi}

$$\vbox{\tabskip=0pt \baselineskip=10pt
\def\tablerule{\noalign{\hrule}}

\halign{&#\vrule&\strut~~$#$~~&#\vrule&\strut$~#~$&#\vrule&\strut$~#~$&#\vrule\cr\tablerule

&S^z&&P&&\hbox{\sc energy eigenvalues}&\cr\tablerule
&6&&~~0&&\u{18}&\cr\tablerule
&5&&~~0&&\u{15}&\cr\tablerule
&~&&\pm2\pi/6&&\u{16}&\cr\tablerule
&~&&\pm4\pi/6&&{18}&\cr\tablerule
&~&&~~\pi&&\u{19}&\cr\tablerule
&4&&~~0&&\twoa&\cr\tablerule
&~&&\pm2\pi/6&&\twob&\cr\tablerule
&~&&\pm4\pi/6&&\twoc&\cr\tablerule
&~&&~~\pi&&\twod&\cr\tablerule
&3&&~~0&&\threea&\cr\tablerule
&~&&\pm2\pi/6&&\threeb&\cr\tablerule
&~&&\pm4\pi/6&&\threec&\cr\tablerule
&~&&~~\pi&&\threed&\cr\tablerule
&2&&~~0&&\foura&\cr\tablerule
&~&&\pm2\pi/6&&\fourb&\cr\tablerule
&~&&\pm4\pi/6&&\fourc&\cr\tablerule
&~&&~~\pi&&\fourd&\cr\tablerule}}$$

\vfill\eject
\centerline{{\bf Table 4} (cont'd.)}

\setbox\strutbox=\hbox{\vrule height12pt depth5pt width0pt}
\def\strut{\relax\ifmmode\copy\strutbox\else\unhcopy\strutbox\fi}

$$\vbox{\tabskip=0pt \baselineskip=10pt
\def\tablerule{\noalign{\hrule}}

\halign{&#\vrule&\strut~~$#$~~&#\vrule&\strut$~#~$&#\vrule&\strut$~#~$&#\vrule\cr\tablerule

&S^z&&P&&\hbox{\sc energy eigenvalues}&\cr\tablerule
&1&&~~0&&\fivea&\cr\tablerule
&~&&\pm2\pi/6&&\fiveb&\cr\tablerule
&~&&\pm4\pi/6&&\fivec&\cr\tablerule
&~&&~~\pi&&\fived&\cr\tablerule
&0&&~~0&&\sixa&\cr\tablerule
&~&&\pm2\pi/6&&\sixb&\cr\tablerule
&~&&\pm4\pi/6&&\sixc&\cr\tablerule
&~&&~~\pi&&\sixd&\cr\tablerule}}$$

\vfill\eject
\end